\documentstyle[aps,prd,graphicx,amsfonts,amssymb]{revtex}
\newcommand{\be}{\begin{equation}}
\newcommand{\ee}{\end{equation}}
\newcommand{\bea}{\begin{eqnarray}}
\newcommand{\eea}{\end{eqnarray}}
\newcommand{\nn}{\nonumber}
\newcommand{\lb}{\label}
\newcommand{\ind}[2]{^{#1}_{\mbox{\scriptsize #2}}}
\newcommand{\al}[2]{\alpha\ind{#1}{#2}}
\newcommand{\nf}{n_{\mbox{\scriptsize f}}}
\newcommand{\mpi}{m_{\pi}}

\begin{document}

\pagestyle{myheadings}

\parindent 0mm
\parskip 6pt

\centerline{\Large{\bf QCD coupling below 1\ GeV from}}
\vspace{0.2truecm}
\centerline{\Large{\bf quarkonium spectrum}}
\vspace{0.5truecm}
\centerline{\bf{M.\ Baldicchi$^{\dagger}$, A.\ V.\ Nesterenko$^{\ast}$,
G.\ M.\ Prosperi$^{\dagger}$, C.\ Simolo$^{\dagger}$}}
\vspace{0.3truecm}
{}\centerline{\small{$^{\dagger}\,$ Dip. di Fisica, Universit\`a
di Milano and INFN, Sezione di Milano }}
\centerline{\small{Via Celoria 16, I20133 Milano, Italy}}
{}\centerline{\small{$^{\ast}\,$ Bogoliubov Laboratory of Theoretical
Physics}}
{}\centerline{\small{
Joint Institute for Nuclear Research, Dubna, 141980, Russia}}


\begin{abstract}
\noindent
In this paper we extend the work synthetically presented in
Ref.~\cite{Baldicchi:2007ic} and give
theoretical details and complete tables of
numerical results.
We exploit calculations within a Bethe-Salpeter (BS)
formalism adjusted for QCD, in order to extract an ``experimental''
strong coupling $\alpha_{\rm s}^{\rm exp}(Q^2)$ below 1~GeV by
comparison with the meson spectrum. The BS potential follows from a
proper ansatz on the Wilson loop to encode confinement and is the sum
of a one-gluon-exchange and a confinement terms. Besides, the common
perturbative strong coupling is replaced by the ghost-free expression
$\alpha_{\rm E}(Q^2)\, $ according to the prescription of Analytic
Perturbation Theory (APT).\\
The agreement of $\alpha_{\rm s}^{\rm exp}(Q^2)$ with the APT
coupling $ \alpha_{\rm E}(Q^2) $ turns out to be reasonable
from 1~GeV down to the 200~MeV scale, thus confirming quantitatively
the validity of the APT prescription. Below this scale, the
experimental points could give a hint on the vanishing of $\alpha_{\rm
s}(Q^2)$ as $Q$ approaches zero. This infrared behaviour would be
consistent with some lattice results and a ``massive'' generalization
of the APT approach.\\
As a main result, we claim that the combined BS-APT theoretical
scheme provides us with a rather satisfactory correlated
understanding of very high and rather low energy phenomena from few
hundreds MeV to few hundreds~GeV.

\end{abstract}


\section{Introduction}
\begin{figure}[htbp!]
 \begin{picture}(420,600)
 \put(20,300){\includegraphics[height=.30\textheight,
                      width=.32\textheight]{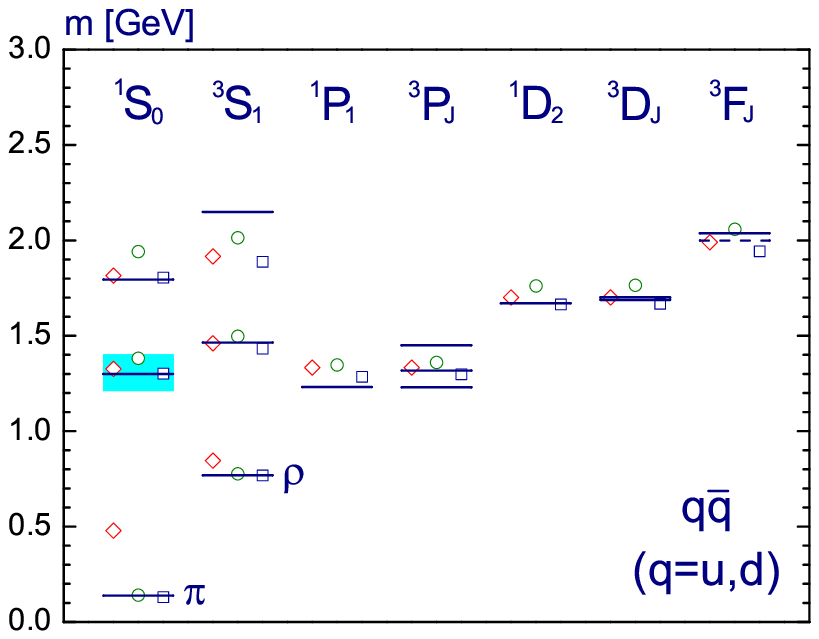}}
 \put(290,300){\includegraphics[height=.30\textheight,
                      width=.32\textheight]{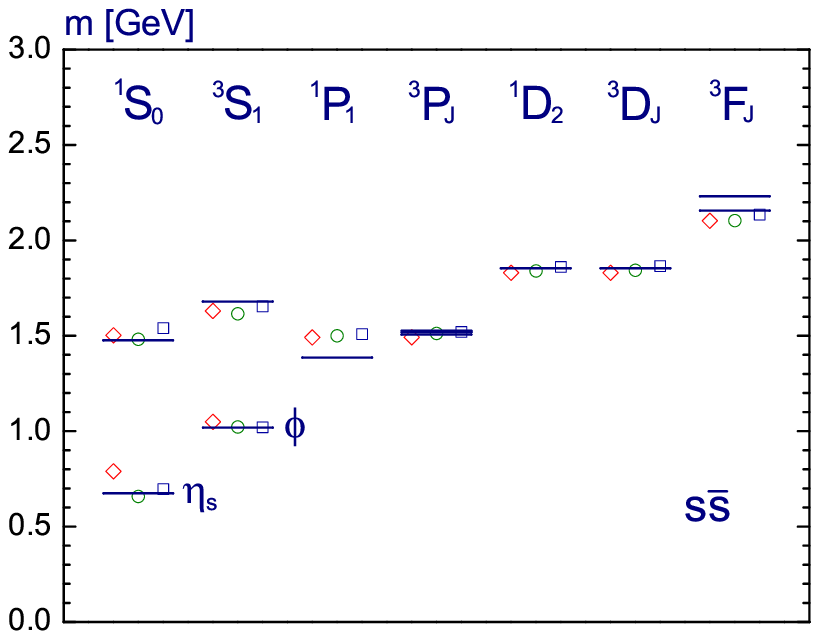}}
 \put(20,50){\includegraphics[height=.30\textheight,
                      width=.32\textheight]{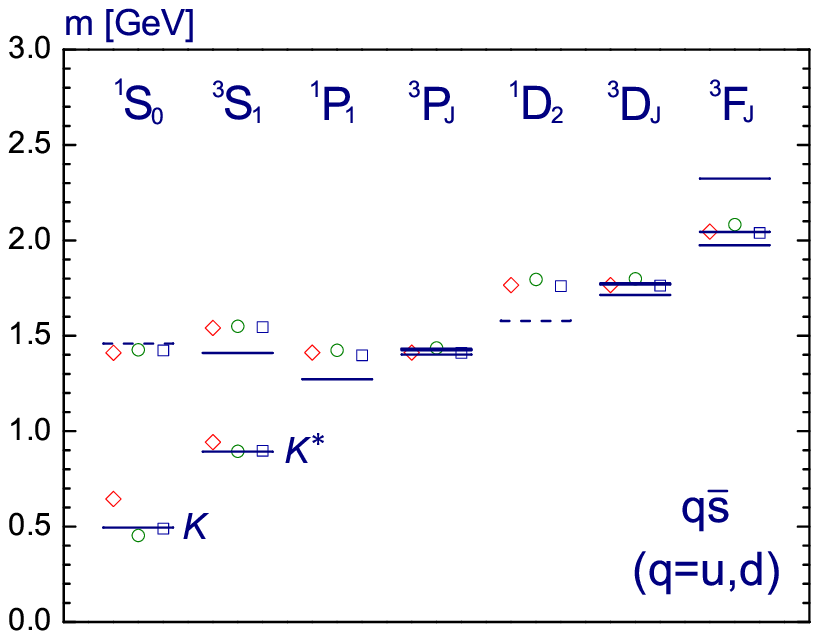}}
 \put(290,50){\includegraphics[height=.30\textheight,
                      width=.32\textheight]{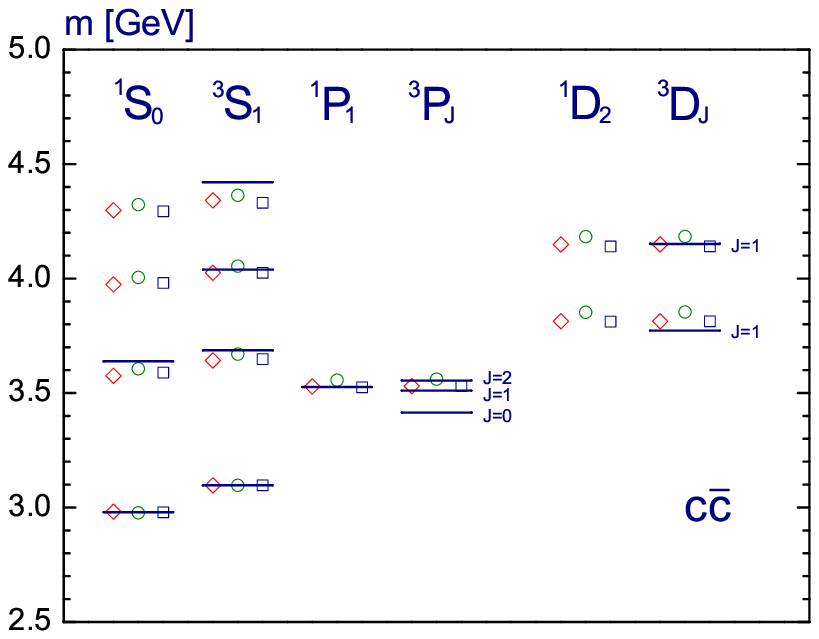}}
 \end{picture}
\caption{
Quarkonium spectrum, three different calculations.
Diamonds refer to the freezing prescription for the running coupling, squares
and circles refer to the calculation with the 1-loop analytic coupling
(\ref{ARC1L}) and two different expressions for running constituent masses
of light quarks, a solution of the Dyson-Schwinger equation and a 
phenomenological function of the c.m. quark momentum respectively. 
Horizontal lines represent experimental data.}
\label{spect}
\end{figure}

This paper is an extended and more thorough analysis of the work
already presented concisely in Ref.~\cite{Baldicchi:2007ic}.
Here we try a self-contained theoretical discussion of the method,
with all the premises and the necessary details. We report the
complete numerical tables of our results, that were originally displayed
only graphically, and we discuss them extensively.
 As it is well known, a rather consistent picture of many strong
 high energy processes was obtained by perturbative QCD, if the
 running coupling
$ \bar{\alpha}_{\rm s}(Q^2)\,$,
as derived from the
 renormalization group equation, is used.
  Inversely, if the QCD scale value $\Lambda_{n_f=5}\simeq 200\;$MeV
 is taken, the values of $\alpha_{\rm s}$ extracted from the data for the
 appropriate $Q$ or $\sqrt{s}\,$ fit rather well the theoretical
$ \bar{\alpha}_{\rm s}(Q^2)\,$
curve, with few exceptions. Quite important, the
 2-loop level of approximation for
$ \bar{\alpha}_{\rm s}\,$
seems to be sufficient
\cite{Bethke:2006ac}
for practical size of data errors.\\
   Unhappily,
$ \bar{\alpha}_{\rm s}(Q^2)$
 develops (at any loop level) unphysical
 singularities for $Q\sim\Lambda_{n_f=3}\sim 400\;$MeV, that
 makes the expression useless in the infrared region. This is a
 particularly serious difficulty in any type of potential model
 in which $Q$ should be identified with the momentum transfer,
 that takes typically values below 1~GeV, according to the state
 and to the mass of the quarks implied.\\
 Among various proposals to eliminate these singularities (see,
 e.g., Section~3 in Ref.~\cite{Prosperi:2006hx}), we mention here
 two particular ones, i.e., the freezing hypothesis, that simply
 consists in assuming that
$ \bar{\alpha}_{\rm s}(Q^2)\,$
 freezes to a certain
 maximal value $H$ in the infrared (IR) region as a
 consequence of non-perturbative effects, and the analytization
 prescription of the APT approach \cite{ShSol96-7}. The latter
 requires 
$ \bar{\alpha}_{\rm s}(Q^2)\,$
 to satisfy a dispersion relation with the only
 unitary cut for $- \infty < Q^2 <0 $ (throughout this paper we
 assume the momentum scale to be spacelike $Q^2 = -q^2 > 0$) and uses
 perturbation theory to evaluate the spectral function.\\
 On the other hand, in the last years various relativistic formalisms
 have been proposed in the context of QCD, or QCD motivated, that
 take confinement into account and evaluate the meson (and baryon)
 spectrum in the light and in the heavy quark sectors. Among the most
 recent works we remind, e.g.,~\cite{Dosch:1987sk,Baker:1994nq,Hecht:2000xa,Jugeau:2003df,Badalian:2007km,Ebert:2005ha} and references therein.
 In this paper we reverse somewhat the point of view, that is, we
 take advantage of the comparison between the calculated meson spectrum
 and the data in order to gain information on the infrared behaviour
 of the QCD coupling that we shall compare with APT. To this end,
we exploit a Bethe-Salpeter (BS) like
 formalism (\emph{second order} BS formalism) developed
 in~\cite{BMP} and applied with a certain success to the calculation of
 a rather complete quarkonium (meson) spectrum in
 Refs.~\cite{BP,Baldicchi:2004wj}.
 The formalism is essentially derived from the QCD
 Lagrangian taking advantage of a Feynman-Schwinger representation
 for the solution of the iterated Dirac equation in an external
 field. Confinement is encoded through an ansatz on the Wilson
 loop correlator; indeed the expression $i\ln W$ is written as the
 sum of a one-gluon exchange~(OGE) and an area term
\begin{equation}
 i\ln W = (i\ln W)_{\rm OGE}+ \sigma S \,.
\label{eq:wilson}
\end{equation}
By means of a three dimensional reduction, the original BS
equation takes the form of the eigenvalue equation for a squared
bound state mass
\begin{equation}
    M^2 = M_0^2 + U_{\rm OGE}+U_{\rm Conf} \,,
\label{eq:M2}
\end{equation}
where $M_0$ is the kinematic term
$ M_0 = w_1+w_2 = \sqrt{m_{1}^2 + {\bf k}^2} + \sqrt{m_{2}^2 + {\bf
     k}^2}$,
${\bf k}$ being the c.m.\ momentum of the quark, $m_1$ and $m_2$
the quark and the antiquark constituent masses, and $U=U_{\rm
OGE}+U_{\rm Conf}$ the resulting potential.\\
As a consequence of ansatz (\ref{eq:wilson}), the perturbative part of
the potential $ U_{\rm OGE}$ turns out to be proportional to
$ \alpha_{\rm s}(Q^{2})\, $, where in a sense $ \alpha_{\rm s}(Q^{2})\, $
should be identified as an effective charge of the type proposed in
\cite{DMW} and denoted in~\cite{Prosperi:2006hx} with
$\alpha_{\rm SGD}(Q^{2})$.\\
Calculations have been performed in Refs.~\cite{BP,Baldicchi:2004wj} by
using both a frozen and the 1-loop analytic coupling
$\alpha^{(1)}_{\rm E}(Q^{2})$
with an effective scaling constant
$\Lambda_{n_f=3}^{(1,{\rm eff})}\simeq 200\,$MeV
(see Eq.\ (\ref{ARC1L}) below), which is equivalent at the 3-loop level to
$\Lambda_{n_f=3}^{(3)}\simeq 400\,$MeV or to the world average
$\Lambda_{n_f=5}^{(3)}\simeq 200\,$MeV.\\
The results of the two sets of calculations are relatively similar
for the heavy-heavy quark states. However, for the $1$S
states involving light and strange quarks, quite different results
have been obtained in the two cases. In the case of a frozen
coupling the $\pi$ and $K$ masses turn out to be too high, independently
of how small the light quark mass is taken (see Fig.~\ref{spect});
e.g., if we fit the light and the strange quark masses to the $\rho$
and the $\phi$ masses, we find $m_\pi\sim 500\,$MeV and $m_K\sim 700\,$MeV,
respectively. On the contrary, if appropriate values for the quark
masses are chosen, the $\pi,\,\,\rho,\,\,K,\,\,K^*,\,\,\phi$
masses can be rather well reproduced when the analytic coupling
$\alpha_{\rm E}^{(1)}(Q^{2})$ is used. This occurrence
strongly supports
the use of the analytic coupling in the BS framework.\\
In this work we undertake a thorough analysis of this issue, by comparing
our theoretical results, obtained for a certain
choice of the parameters and the analytic coupling, with the
results of a similar calculation performed by means of a fixed value
of~$\alpha_{\rm s}$, for every quark-antiquark state.
We denote by $\alpha_{\rm s}^{\rm th}$ the value that reproduces the
same theoretical result as obtained with
$ \alpha_{\rm E}^{(1)}(Q^2)$ and
by $\alpha_{\rm s}^{\rm exp}$ the corresponding value that reproduces
the experimental mass. The value $\alpha_{\rm s}^{\rm th}$ is
then used to identify an effective $Q$ pertaining to that
particular state, which is to be understood as the argument of
the related ``experimental'' coupling $\alpha_{\rm s}^{\rm exp}(Q^2)$.\\
Since only the leading perturbative contribution in the BS kernel has
been included, a rough estimate of NLO effects on the
$\alpha_{\rm s}^{\rm exp}$ value leads to a relative theoretical error
which spans from $20\%$ to much less than $1\%$ throughout the
spectrum according to the quark masses involved. Furthermore,
since coupling among different quark-antiquark channels has not been
taken into account, the theoretical masses are expected to reproduce
the experimental ones within the half width $\Gamma/2$ of the state.
In the framework of the BS formalism these are the most relevant sources
of theoretical error and overwhelm all other errors, like those
related to the three dimensional reduction. When relevant, the experimental
error, related to the uncertainty of the experimental mass is added to the
theoretical one.\\
It should be noted that our results are model dependent in the sense
that they do depend on the ansatz~(\ref{eq:wilson}). More sophisticated
models also exist, like the Stochastic vacuum model~\cite{Dosch:1987sk}
and dual QCD~\cite{Baker:1994nq}, but they turn out to be
considerably difficult to implement in the BS formalism.\\
If compared with the 3-loop analytic curve
$\alpha_{\rm E}^{(3)}(Q^{2})$ normalized at the world average
$\alpha_{\rm E}^{(3)}(M_Z^2)=0.1176\,$ ($\Lambda_{n_f=5}^{(3)}=236\,$MeV),
the values of $\alpha_{\rm s}^{\rm exp}(Q^2)$ fit it
rather well within error bars in all the region
$\Lambda_{n_f=3}^{(3)}/2< Q <3\Lambda_{n_f=3}^{(3)}\,$ (being
$\Lambda_{n_f=3}^{(3)}= 417\,$MeV according to
threshold matching). In the region
$Q<\Lambda_{n_f=3}^{(3)}/2\,$ our data fall below
$ \alpha_{\rm E}^{(3)}(Q^2)\,$, and, with the limitation due to the
large errors, this could be interpreted
as a hint on the vanishing of $\alpha_{\rm s}(Q^2)$ as $Q\to0\,$, or 
on the existence of
a finite limit lower than the universal Shirkov-Solovtsov freezing
value (see e.g.~\cite{Badalian:2007km}). Note, however, that, as it 
will be discussed extensively in Sec.\ 5,
the experimental situation is particularly uncertain in this region and the
theoretical treatment problematic. Nonetheless, it is worthwhile to mention
that the former behaviour is consistent with a recently
developed ``massive'' version of the analytic approach for the QCD
coupling~\cite{MAPT} (see Sec.\ 2), and 
some results from lattice simulations.\\
Finally let us stress that the
choice to compare $\alpha_{\rm s}^{\rm exp}(Q^2)\,$
with the 3-loop expression
$\alpha^{(3)}_{\rm E}(Q^2)$  was to stay as close as possible to the usual
practice in perturbation theory. In APT, however, when the appropriate
small change of scale is made ($\Lambda_{n_f=5}^{(2)}\simeq 258$ MeV rather
than $\Lambda_{n_f=5}^{(3)}\simeq 236$ MeV if both normalized at the
$Z$ mass\footnote{Note that these values of the scale constant
turn out to be somewhat larger than the perturbative values, as given e.g., by
Bethke~\cite{Bethke:2006ac}, with the same normalization for
$\alpha_{\rm s}\,$.})
the 2-loop coupling $\alpha^{(2)}_{\rm E}(Q^2)$ is practically
indistinguishable
from $\alpha^{(3)}_{\rm E}(Q^2)$ in the entire
Euclidean range and can be used instead of the latter for all
practical purposes.\\
The layout of the paper is as follows. In Sec.~2
the ghost-pole problem is discussed
and an overview of the key points of analytic approach
to~QCD is given.
In Sec.~3 an explicit expression for the BS potential
$U$ is given
and the mathematical method to treat the eigenvalue
problem for the squared mass operator $M^2$
is described. Sec.~4
is devoted to the strategy for extracting $\alpha_{\rm s}^{\rm exp}(Q^2)$
from the data and to errors estimate.
Finally, in Sec.~5
our results are discussed and a match of the 
extracted QCD coupling with relevant high energy experimental
data is attempted via 
Analytic Perturbation Theory.

Some technical material is exposed in Appendices. A brief review of
the derivation of the second order BS formalism and of the expression
of $M^2$ from the ansatz~(\ref{eq:wilson}) is given in App.~A.
Numerical tables in App.~B display all results in details. In App.~C
an useful formula for 3-loop analytic coupling (the spectral density)
is explicitly given and compared with the usual 3-loop perturbative
coupling.


\section{Analytic approach to QCD}

The renormalization group~(RG) method is an inherent part of theoretical
description of strong interaction processes. It is usually employed to
improve the results of perturbation theory in the high energy region.
However, a straightforward application of the RG method to perturbative
expansion eventually gives rise to unphysical singularities of both the
RG-invariant coupling function\footnote{For example, the one-loop strong
running coupling~(\ref{A1LPert}) possesses the so-called Landau (or
spurious) pole in the low energy region. This problem can not be solved by
the inclusion of higher loop corrections since the latter just give rise
to additional singularities of the cut type.} and physical observables.
The presence of these singularities contradicts the general principles of
the local QFT\footnote{It is worth noting also that the results of lattice
simulation testify to the absence of spurious singularities of the QCD
coupling at low energies, see, e.g., a recent overview in Sec.~2 of
paper~\cite{dvFourier03} as well as original papers~\cite{Lattice}.} and
severely complicates theoretical analysis of hadron dynamics in the IR
domain.

There is a variety of the nonperturbative approaches to handle the
strong interaction processes at low energies (for a recent review of
this issue see Sec.~3 in paper~\cite{Prosperi:2006hx} and references
therein). Some of such methods originate in the general properties of
perturbative power series for the QCD observables in the framework of
the RG formalism. For instance, these are the method of effective
charges~\cite{EffChar}, the ``optimal conformal mapping''
method~\cite{Fischer} (see also Ref.~\cite{Maxwell}), the ``optimized
perturbation theory''~\cite{OPT}, and some others. There is also a
number of approaches which impose nonperturbative constraints either
on the strong running coupling (see, e.g.,
Refs.~\cite{Cvetic,AlekArbu}) or on the RG $\beta$~function (see,
e.g., Refs.~\cite{PRD,Review,Raczka}). In this paper we will exploit
the so-called Analytic Perturbation Theory (APT)~\cite{ShSol96-7} and
its recent ``massive'' modification~\cite{MAPT}, which are briefly
overviewed in what follows.

The analytic approach to QFT constitutes the next step (after the
RG-summation) in improving the perturbative results. Specifically, in
addition to the property of renormalizability this method retains a
general feature of local QFT, the property of causality. The analytic
approach has first been devised in the context of Quantum
Electrodynamics~\cite{AQED}, and then extended to the QCD case about ten
years ago~\cite{ShSol96-7}. The basic merits of the analytic approach to
QCD are the absence of unphysical singularities of the invariant charge
and the enhanced stability of outcoming results with respect to both
higher loop corrections and choice of renormalization
scheme~\cite{ShSol98-9}. Besides, this method enables one to process the
spacelike and timelike data in a congruent way~\cite{MS:97}. A fresh
review of the analytic approach to QCD and its applications can be found
in paper~\cite{APT-07} (a generalization of APT for fractional powers of
$\alpha_{\rm s}$ was implemented in Ref.~\cite{Bakulev:2006ex}).

Usually, in the framework of RG--improved perturbation theory a QCD
observable $D(Q^2)$ of a single argument $Q^2= -q^2 \ge 0$ (spacelike
momentum transfer squared) can be represented as power series in the
strong coupling~$\bar{\alpha}_{\rm s}(Q^2)$:
\begin{equation}
\label{DPert}
D\ind{}{\tiny{PT}}(Q^2) = 1 + \sum_{n \ge 1}
d_{n}\left[ \bar{\alpha}_{\rm s}(Q^2) \right]^{n},
\end{equation}
where $d_n$ are the relevant perturbative coefficients. However, in
the IR domain this expansion becomes inapplicable due to spurious
singularities of the running coupling~$\bar{\alpha}_{\rm s}(Q^2)$.
For example, the one-loop expression
\begin{equation}
\label{A1LPert}
\bar{\alpha}_{\rm s}^{(1)}(Q^2)=
\frac{1}{\beta_0}\,\frac{1}{\ln(Q^2/\Lambda^2)}\,,
\qquad
\beta_0 = \frac{1}{4\pi}\left(11 - \frac{2}{3} n_{f}\right)
\end{equation}
possesses both the physical cut along the negative real semiaxis
$Q^2\le 0$ and unphysical pole at~$Q^2=\Lambda^2$.

In the framework of the APT, the power series~(\ref{DPert}) for an
``Euclidean" observable is replaced~\cite{dv99} by the nonpower
expansion
\begin{equation}
\label{DAPT}
D\ind{}{\tiny{APT}}(Q^2) = 1 + \sum_{n \ge 1}
d_{n}\,{\cal A}_n(Q^2)
\end{equation}
over the set of functions
\begin{equation}
\label{AAPT}
{\cal A}_{n}(Q^2) = \int\limits_{0}^{\infty}
\frac{\rho_{n}(\sigma)}{\sigma + Q^2}\, d\sigma.
\end{equation}
Here, the spectral function $\rho(\sigma)$ is defined as the discontinuity
of the relevant power of the perturbative running
coupling~$\bar{\alpha}_{\rm s}(Q^2)$ across the physical cut, namely
\begin{equation}
\label{RhoDef}
\rho_{n}(\sigma) = \frac{1}{\pi}\,\mbox{Im}\!
\left[\bar{\alpha}_{\rm s}(-\sigma - i \varepsilon)\right]^{n}.
\end{equation}
The APT representation for a QCD observable $D(Q^2)$~(\ref{DAPT}) is
free of spurious singularities. Besides, it displays a better
stability (in comparison with the perturbative
parameterization~(\ref{DPert})) with respect to both, higher loop
corrections and choice of the renormalization scheme, see
Ref.~\cite{APT-07} for details. The first-order function
${\cal A}_1(Q^2)$~(\ref{AAPT}) plays the role of the effective
Euclidean QCD coupling at respective loop level:
\begin{equation}
\label{ARC}
\alpha_{\rm E}(Q^2) \equiv {\cal A}_{1}(Q^2) =
\int\limits_{0}^{\infty}
\frac{\rho_{1}(\sigma)}{\sigma + Q^2}\,d\sigma.
\end{equation}
In the one-loop case this equation can be integrated
explicitly~\cite{ShSol96-7}
\begin{equation}
\label{ARC1L}
\alpha_{\rm E}^{(1)}(Q^2) = \frac{1}{\beta_{0}}\!\left[\frac{1}
{\ln(Q^2/\Lambda^2)}+\frac{\Lambda^2}{\Lambda^2-Q^2}\right]\!.
\end{equation}
At the higher loop levels the spectral functions~(\ref{RhoDef})
become rather involved (see App.~C). An extensive numerical study of
the analytic running coupling~(\ref{ARC}) and its ``effective
powers''~(\ref{AAPT}) at various loop levels can be found in
Ref.~\cite{Kurashev:2003pt}. Besides, for practical applications one
can also use simple explicit expressions~\cite{APTApprox} which
approximate the APT functions~(\ref{AAPT}) within reasonable
accuracy.

In order to handle the QCD observables which do not satisfy the
integral representation of the form of Eq.~(\ref{AAPT}), the APT has
to be modified appropriately. For example, the Adler function, being
defined as the logarithmic derivative of the hadronic vacuum
polarization function, satisfies the dispersion relation~\cite{Adler}
\begin{equation}
\label{DDef}
D(Q^2) = Q^2 \int\limits_{4\mpi^2}^{\infty}
\frac{R(s)}{(s + Q^2)^2} \, d s,
\end{equation}
where $R(s)$ denotes the Drell ratio of the electron--positron
annihilation into hadrons. Thus, the Adler function~(\ref{DDef}) can
be expanded over the set of functions
${\cal A}_{n}(Q^2)$~(\ref{AAPT}) only in the limit of the massless
pion $\mpi=0$, since otherwise the analytic properties of
$D(Q^2)$~(\ref{DDef}) in $Q^2$~variable differ from those of
${\cal A}_{n}(Q^2)$~(\ref{AAPT}).

The effects due to the nonvanishing pion mass have been incorporated
into the analytic approach to QCD in Ref.~\cite{MAPT}. In the
framework of the latter formalism the Adler function~(\ref{DDef}) can
be expanded
\begin{equation}
\label{DMAPT}
D\ind{}{{\tiny MAPT}}(Q^2, \mpi^2) = \frac{Q^2}{Q^2 +
4\mpi^2} + \sum_{n \ge 1} d_{n}\, \textsf{A}_{n}(Q^2, \mpi^2)
\end{equation}
over the set of the ``massive'' functions
\begin{equation}
\label{AMAPT}
\textsf{A}_{n}(Q^2, m^2) = \frac{Q^2}{Q^2 + 4 m^2}
\int\limits_{4 m^2}^{\infty} \rho_{n}(\sigma)\, \frac{\sigma -
4 m^2}{\sigma + Q^2}\, \frac{d\sigma}{\sigma}.
\end{equation}
Obviously, in the massless limit Eqs.~(\ref{DMAPT}) and~(\ref{AMAPT})
coincide with the expressions~(\ref{DAPT}) and~(\ref{AAPT}),
respectively. Similarly to the case of the APT~(\ref{ARC}), the
first-order function $\textsf{A}_{1}(Q^2, m^2)$~(\ref{AMAPT}) plays
the role of an effective ``massive'' running coupling at the relevant
loop level, namely
\begin{equation}
\label{MARC}
\alpha(Q^2, m^2) \equiv \textsf{A}_{1}(Q^2, m^2) =
\frac{Q^2}{Q^2 + 4 m^2} \int\limits_{4 m^2}^{\infty}
\rho_{1}(\sigma) \, \frac{\sigma - 4 m^2}{\sigma + Q^2}\,
\frac{d\sigma}{\sigma}.
\end{equation}
It is worthwhile to note that irrespective of the loop level this
coupling possesses the universal IR limiting value $\alpha(Q^2, m^2)
\to 0$ at $Q^2=0$, see Ref.~\cite{MAPT} for details.


\section{BS-model for $q\bar q$ spectrum}

As mentioned, in~\cite{BP,Baldicchi:2004wj} the meson spectrum is
obtained by solving the eigenvalue equation for the squared mass
operator~(\ref{eq:M2}), where the perturbative and confinement part
of the potential are respectively
\bea
& &\langle {\bf k} \vert U_{\rm OGE}  \vert {\bf k}^\prime \rangle=\nn\\
\nn\\
& &  {4\over 3} {\alpha_{\rm s}({\bf Q}^2) \over \pi^2}
     \sqrt{(w_1+w_2) (w_1^\prime + w_2^\prime)
     \over w_1 w_2 w_1^\prime w_2^\prime}
     \Bigg[ - \frac{1}{{\bf Q}^2}
     \bigg( q_{10} q_{20} + {\bf q}^2 - { ( {\bf Q}\cdot {\bf q})^2 \over
{\bf Q}^2 } \bigg) \nonumber \\
& & + {i\over 2 {\bf Q}^2} {\bf k}\times {\bf k}^\prime \cdot ({\bf
     \sigma}_1 + {\bf \sigma}_2 ) + {1\over 2 {\bf Q}^2 } [ q_{20}
(\alpha_1 \cdot {\bf Q}) - q_{10} (\alpha_2\cdot {\bf Q}) ]+
    \nonumber \\
& & + {1\over 6} {\bf \sigma}_1 \cdot {\bf \sigma}_2 + {1\over 4}
    \left ( {1\over 3} {\bf \sigma}_1 \cdot {\bf \sigma}_2 -
    { ( {\bf Q}\cdot \sigma_1)
    ( {\bf Q}\cdot {\bf \sigma}_2) \over {\bf Q}^2 } \right )
    + {1\over 4 {\bf Q}^2 } ( \alpha_1 \cdot {\bf Q}) ( \alpha_2 \cdot
{\bf Q}) \Bigg ]\qquad
\lb{upt}
\eea
and
\be
\langle {\bf k} \vert
U_{\rm Conf}  \vert {\bf k}^\prime \rangle ={\sigma\over ( 2 \pi)^3}
\sqrt{(w_1+w_2) (w_1^\prime + w_2^\prime)
     \over w_1 w_2 w_1^\prime w_2^\prime} \int d^3{\bf r}\,
     e^{i {\bf Q}\cdot{\bf r}}
J^{\rm inst}({\bf r}, {\bf q}, q_{10}, q_{20})
\lb{ucf}
\ee
with
\bea
&& J^{\rm inst}({\bf r}, {\bf q}, q_{10}, q_{20})= { r \over
q_{10}+q_{20}}
   \left[ q_{20}^2 \sqrt{q_{10}^2-{\bf q}^2_\perp} +
q_{10}^2 \sqrt{q_{20}^2 - {\bf q}_\perp^2}\right. +\qquad\nonumber\\
& & \qquad\left. + {q_{10}^2 q_{20}^2 \over \vert
{\bf q}_\perp \vert}
   \left(\arcsin{\vert{\bf q}_\perp \vert \over q_{10} }
   + \arcsin{\vert {\bf q}_\perp\vert \over q_{20}}\right)\right]
   \nonumber\\
& & \qquad - {1\over r}  \left[ {q_{20} \over
    \sqrt{q_{10}^2-{\bf q}^2_\perp}}
   ( {\bf r} \times {\bf q}\cdot \sigma_1 + i q_{10} ({\bf r}\cdot
\alpha_1))\right. \nonumber \\
& & \qquad\left.+ {q_{10} \over \sqrt{q_{20}^2 -
{\bf q}^2_\perp}}
   ( {\bf r}\times {\bf q} \cdot
   \sigma_2 - i q_{20} ( {\bf r}\cdot{\bf \alpha}_2)) \right]\,.
   \label{eq:uconf1}
\eea
Here $ \alpha_j^k $ denote the usual Dirac matrices $\gamma_j^0
\gamma_j^k$, $\sigma_j^k$ the $4 \times 4$ Pauli matrices $ \left( \matrix
{\sigma_j^k & 0 \cr 0 & \sigma_j^k}\right ) $ and
${\bf q} ={ {\bf k}+ {\bf k}^\prime \over 2}\,, \quad {\bf Q}= {\bf k} -
{\bf k}^\prime \,, \quad q_{j0}= {w_j+w_j^\prime \over 2}$, $m_{1}$ and
$m_{2}$ are constituent masses.\\
Eqs.~(\ref{upt}-\ref{eq:uconf1}) follow from the ansatz (\ref{eq:wilson})
and a 3-dimensional reduction of our Bethe-Salpeter equation (see
Refs.~\cite{BMP,BP,Baldicchi:2004wj} and App.~A for the details). Actually
in the calculation of \cite{BP,Baldicchi:2004wj} only the center of
gravity (c.o.g.) masses of the fine multiplets were considered as a rule,
and the spin dependent terms in (\ref{upt}-\ref{eq:uconf1}) (spin-orbit
and tensorial terms) were neglected with the exception of the hyperfine
separation term in (\ref{upt}) proportional to ${1\over 6} {\bf \sigma}_1
\cdot {\bf \sigma}_2\,$. Within this limitation a generally good
reproduction of the spectrum was obtained for appropriate values of the
parameters, as apparent from Fig.~\ref{spect} re-elaborated from
\cite{Baldicchi:2004wj}. Here the results of three sets of calculations
are displayed. Diamonds refer to the usual perturbative 1-loop coupling
(to be replaced in Eq.\ (\ref{upt})), frozen at a maximum value $H\,$,
which has been taken as an additional adjustable parameter. Squares and
circles both refer to the 1-loop APT coupling~(\ref{ARC1L}) with
$\Lambda\simeq 200\,$MeV. For light quarks a running constituent mass was
used too.\footnote{Circles refer to a phenomenological running mass as
function of the c.m.\ quark momentum $m_u^2=m_d^2 = 0.17 |{\bf k}| - 0.025
|{\bf k}|^2 + 0.15 |{\bf k}|^4$. Squares refer to a running constituent
mass resulting from a solution of the DS equation with an analytic RG
running current mass (see App.~A for the details) which was, however, more
in line with an attempt to define an analytic $\alpha_{\rm
E}^{(1)}(Q^{2})$ singular at $Q^2\to 0$ and including
confinement~\cite{PRD,Review} than to~(\ref{ARC1L}).}

\begin{figure}[th]
\centerline{\includegraphics[width=125mm]{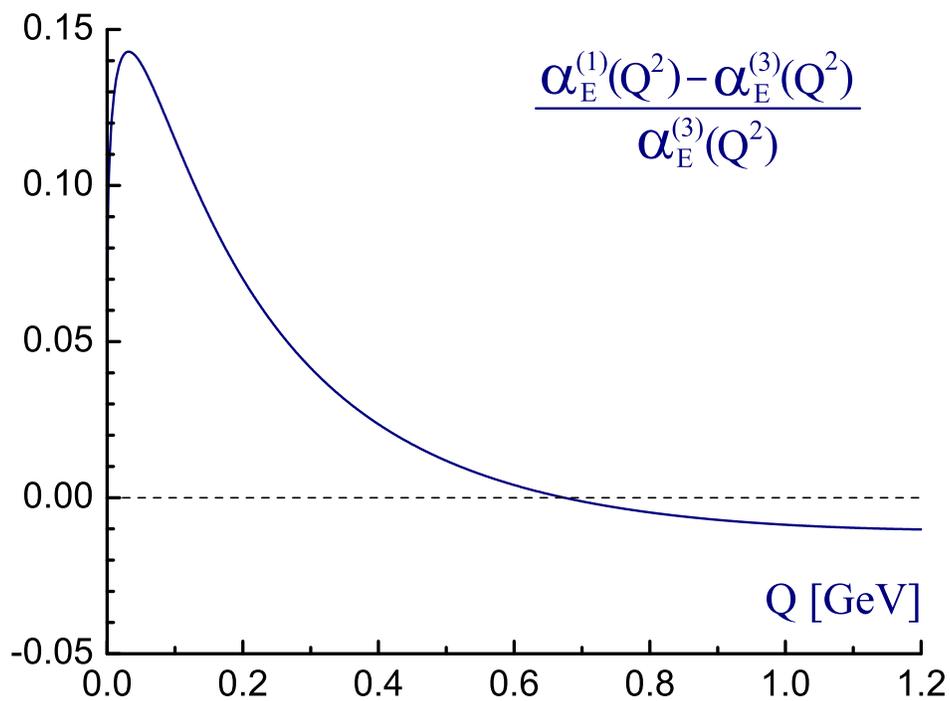}}
\vskip7.5mm
\caption{Relative difference between the one-loop analytic running
coupling $\alpha^{(1)}_{\rm E}(Q^2)$ with $\Lambda^{(1,{\rm eff})}_{n_f=3}
= 193\,$MeV and three-loop $\alpha^{(3)}_{\rm E}(Q^2)$ with
$\Lambda^{(3)}_{n_f=3}=417\,$MeV in the range $0<Q<1.2\,$ GeV.}
\label{dif}
\end{figure}

We stress that only with the choice (\ref{ARC1L}) the $1^1S_0$
state has been correctly reproduced when light and strange quarks were
involved, as in the case of $\pi$ and the $K$ mesons.\\
In the present work a similar calculation with the input
(\ref{ARC1L}) and a slight different choice of the parameters
is made (preliminary results were given in~\cite{Baldicchi:2006az}).
First, we have fixed the string tension to the value $\sigma=0.18\,\,{\rm
GeV^2}\,$ (consistent with other phenomenology and lattice
simulations) and the scale constant to
$\Lambda_{n_f=3}^{(1,{\rm eff})}=193\,$MeV. The whole set of remaining
parameters, all the quark masses, are then determined by fitting
the $\pi\,$, $\phi\,$, $J/\psi\,$ and $\Upsilon\,$
masses. It turns out $m_u=m_d=196\,$MeV, $m_s=352\,$MeV,
$m_c=1.516\,$GeV and $m_b=4.854\,$GeV. The results for the meson
spectrum are given in the fourth column of tables in App.~B.\\
The chosen value for $\Lambda_{n_f=3}^{(1,{\rm eff})}$ has been
dictated by the comparison with the 3-loop analytic coupling
normalized at the Z boson mass (see App.~C) according to the
world average. As displayed in Fig.~\ref{dif} the relative difference
between the two curves is no more than 1$\%$ in the region
$0.5<Q<1.2\,$GeV, to which the states used as an input
in the calculation belong.\\
Furthermore, as already noted, our equations refer to a single definite
quark-antiquark channels. So, having correct relativistic
kinematics, they do not include coupling with
other channels like any potential model (see App.~A). Then we can not
expect to have any insight into the splitting of
over-threshold complicated multiplets which involve mixture of different
states. Even the position of the c.o.g.\ mass is expected to be reproduced
only within one-half of the width of the state. This has been taken into
account in the estimate of the theoretical error (see Sec.~4).\\
The resolution method of the eigenvalue equation for the operator
(\ref{eq:M2}, \ref{upt}-\ref{eq:uconf1}) we have used in
\cite{BP,Baldicchi:2004wj} and in the present work can be summarized
in the following way.

a) In the static limit the problem can be reduced to the corresponding
one for the center of mass Hamiltonian (see App.~A)
\be
H_{\rm CM} = w_1 + w_2  - {4 \over 3} {\alpha_{\rm s} \over r } +  \sigma
r \, .
\label{eq:static}
\ee

b) The eigenvalue equation for (\ref{eq:static}) is solved
for a convenient fixed $\alpha_{\rm s}$ by the
Rayleigh-Ritz method, using the three dimensional harmonic oscillator
basis and diagonalizing a $30 \times 30$ matrix.

c) The square of the meson mass is evaluated as $\langle \phi_a |
M^2 |\phi_a\rangle $, $ \phi_a $ being the eigenfunction obtained
in step b) (with $a$ the whole set of quantum numbers) and the
operator $M^2$ given by Eq.\ (\ref{eq:M2}).

d) Prescription c) is equivalent to treat $ M^2-H_{\rm CM}^2 $ as a first
order perturbation. Consistently the hyperfine separation should be given
by
\begin{eqnarray}
& & (^3m_{nl})^2-(^1m_{nl})^2 =
 {32 \over 9\pi} \int_0^\infty \!
     dk \, k^2  \int_0^\infty \! dk^{\prime} \,
     k^{\prime 2} \varphi_{nl}^* (k)  \varphi_{nl} (k^{\prime})\,\times
\nonumber \\
& &  \qquad \qquad \qquad \sqrt {{w_1+w_2 \over w_1 w_2}}
     \sqrt {{w_1^{\prime}+w_2^{\prime}
     \over w_1^{\prime} w_2^{\prime}}} \int_{-1}^1 \! d\xi \,
     \alpha_{\rm s}({\bf Q}^2) P_l(\xi)\, ,
\label{eq:hyper}
\end{eqnarray}
where $\varphi_{nl}$ is the radial part of the complete eigenfunction
$\phi_a\,$. However, in the case of the states involving light and strange
quarks the quantity is
further corrected to the second order of perturbation theory.\\
For the quark masses and string tension $\sigma$ in~(\ref{eq:static})
we have used the same values listed above and for what regards
$\alpha_{\rm s}\,$, that is supposed to be a constant
in~(\ref{eq:static}), we have taken $\alpha_{\rm s}=0.35\,$, which is
the typical value used in non-relativistic calculations and also the
freezing value adopted in~\cite{BP}.


\section{Extracting $\alpha_{\rm s}^{\rm exp}(Q^2)$ from the data}

One focus now on the reversed problem, i.e., the determination
of the $\alpha_{\rm s}^{\rm exp}(Q^2)$ values at the characteristic
scales of a selected number of ground and excited states.\\
In order to estimate $\alpha_{\rm s}^{\rm exp}(Q^2)$ at low scales
one needs first to assign an effective $Q$-value to each state.
To this end one first rewrites the squared mass, as given by point c)
in Sec.\ 3, more explicitly as the sum of the unperturbed part, the
perturbative and the confinement one respectively

\be
m^2_{a}
=\langle\phi_a|M_0^2|\phi_a\rangle+
\langle\phi_a|U_{\rm OGE}|\phi_a\rangle+
\langle\phi_a|U_{\rm Conf}|\phi_a\rangle\,.
\lb{m_th}
\ee

Here
$U_{\rm OGE}$ is given by the second line of (\ref{upt}) and $U_{\rm Conf}$
by Eq.\ (\ref{ucf}) and first two lines of (\ref{eq:uconf1}).
From the OGE contribution we then extract for each state the
fixed coupling value $\alpha_{a}^{\rm th}\,$ which leads
to the same theoretical mass as by using $\alpha_{\rm E}^{(1)}(Q^2)$
given by Eq.\ (\ref{ARC1L}). This can be done by means of the
relation
\be
\langle\phi_a|U_{\rm OGE}|\phi_a\rangle
\equiv\langle\phi_a|\alpha_{\rm E}^{(1)}({\bf Q} ^2){\cal O}
({\bf  q};{\bf Q})
|\phi_a\rangle=\alpha_{a}^{\rm th}\langle\phi_a|{\cal O}
({\bf q};{\bf Q})
|\phi_a\rangle,
\lb{a_th}
\ee

where ${\cal O}({\bf q};{\bf Q})$ can be drawn again by the
second line of Eq.~(\ref{upt}). The effective momentum transfer $Q_a$
associated to each bound state is then identified by equating

\be
\alpha_{\rm E}^{(1)}(Q_a^2)=\alpha_{a}^{\rm th}\,.
\lb{Qeff}
\ee

The next step is to search for the correct (fixed) value of the
coupling that exactly reproduces the experimental mass of
each state. This is defined by the relation

\be
\langle\phi_a|M_0^2|\phi_a\rangle+
\alpha_{\rm s}^{\rm exp}(Q^2_a)\langle\phi_a|{\cal O}({\bf q};{\bf Q})
|\phi_a\rangle +
\langle\phi_a|U_{\rm Conf}|\phi_a\rangle=m^2_{\rm exp}\,,
\lb{m_exp}
\ee

so that, by combining Eqs.\ (\ref{m_th}), (\ref{a_th}) and (\ref{m_exp})
we finally obtain

\be
\alpha_{\rm s}^{\rm exp}(Q^2_a)=\frac{m^2_{\rm exp}-m^2_{a}+
\alpha_{a}^{\rm th}\langle\phi_a|{\cal O}({\bf q};{\bf Q})
|\phi_a\rangle}{\langle\phi_a|{\cal O}({\bf q};{\bf Q})|\phi_a\rangle}\,.
\lb{a_exp}
\ee

This procedure has been applied to a number of light-light,
light-heavy and heavy-heavy ground as well as excited states.\\
Note that, apart from the particular ansatz made in~(\ref{eq:wilson})
to take into account confinement, the other relevant approximations
are:
\\
\\
i) only the leading perturbative contribution is included
in~(\ref{eq:wilson}) and so in the potential;
\\
ii) quark antiquark annihilations and couplings with other
channels have been ignored;
\\
iii) an instantaneous approximation is involved in deriving the
eigenvalue equation for~(\ref{eq:M2}) from the original BS equation.
\\
\\
As even the experience with QED suggests, retardation corrections
are expected to be relevant for the hyperfine and possibly fine
splitting, but of minor importance for the positions of the c.o.g.\
of the multiplets that we essentially use
to evaluate $\alpha_{\rm s}^{\rm exp}(Q^2)$. \\
Thus, as we told, the main sources of theoretical error in the
whole procedure are expected to arise from neglecting the NLO
contribution to the BS kernel as well as the coupling with other
channels. For what concerns the former (point i)), it is worth
noting that the next to leading contribution to the perturbative
part of the BS-kernel comes from four diagrams with two-gluon
exchange; two triangular graphs containing a four-line vertex of
the type $g^2\phi^*\phi A_{\mu}A^{\mu}$ and two three-line
vertices $g\phi^*\partial_{\mu}\phi A^{\mu}$ (the spin independent
part of our second order formalism is quite similar to scalar
QED), one {\it fish diagram} with two four-line vertices, and a
crossing box with four three line vertices. If the renormalization
scale is identified with the momentum transfer $Q$ the fish graphs
contribution is completely reabsorbed in the renormalization. On
the other hand, a somewhat crude estimate of the contribution of
each of the two triangular graphs gives
\begin{equation}
        I_{\rm triang} \sim 4\left({4 \over 3}\,
        \alpha_{\rm s}\right)^2 {9m^2 \over 4Q^2 + 2m^2}
\end{equation}
and for the crossing box graph, similarly
\begin{equation}
     I_{\rm crsbox} \sim  {64 \over 3}
     \left({4 \over 3}\,\alpha_{\rm s}\right)^2
{m^4 \over (Q^2 + m^2 + k^2)^2}.
\end{equation}
These expressions have to be compared with the leading one-gluon
term we have used (see Eq.~(\ref{eq:imom}) of App.~A)
\begin{equation}
      I_{\rm OGE} \sim 16\pi {4 \over 3 }\,\alpha_{\rm s} {m^2
          \over Q^2}\,.
\end{equation}
Putting all things together, the overall error due to the omission
of such NLO contributions to the BS kernel is then
\begin{equation}
{\Delta I\over I} = \sqrt{\left(2\,{I_{\rm triang}\over I_{\rm
OGE}}\right)^2+\left({I_{\rm crsbox}\over I_{\rm
OGE}}\right)^2}\,,
\end{equation}
and this produces
\begin{equation}
{{\Delta\cal O}\over{\cal O}}\sim{\Delta I\over I}\,.
\end{equation}
By using Eqs.~(\ref{a_th}-\ref{a_exp}), after some algebra it is easy
to recognize that the NLO effects on $\alpha_{\rm s}^{\rm exp}$ turn
out to be of the same order, that is
\begin{equation}
\Delta_{\rm NLO}\alpha_{\rm s}\sim\alpha_{a}^{\rm th}\,{\Delta
I\over I}\,,
\end{equation}
which is what is assumed in the foregoing.
The NLO errors do not exceed $5\%$ for heavy
quark states while they are enhanced up to $20\%$ when light and
strange quarks are
involved.\\
Finally, since the strength of the neglected coupling with other
channels (OC) is obviously measured  by the width $\Gamma_a$ of
the state, one roughly estimates an error of the order of $\Delta
m_a\sim \Gamma_a/2\,$ in the evaluation of $m_a\,$. On this
ground, for each determination of
$\alpha_{\rm s}^{\rm exp}\,(Q^2_a)$
the related theoretical error is given by
\begin{equation}
\Delta_{\rm \Gamma}\alpha_{\rm s}=\frac{m_a}
{\langle\phi_a|{\cal O}({\bf q};{\bf Q})|\phi_a\rangle}\,\Gamma_a\,.
\label{err}
\end{equation}
Usually the error $\Delta m_{\rm exp}$ on the experimental mass $ m_{\rm exp}$
is much smaller than $\Gamma_a/2\,$. When, however, this is not the case
one has to consider also the experimental error
$\Delta_{\rm exp}\alpha_{\rm s}\,$, obtained from (\ref{err}) by
replacing $m_a\,\Gamma_a$ with $2m_{\rm exp}\Delta m_{\rm exp}\,$.

Before discussing our results some comments are in order.\\
First note that in the evaluation of $Q_a$ in (\ref{a_th}) one has
neglected the hyperfine splitting which however was taken
into account in (\ref{a_exp}), bringing possibly to different values of
$\alpha_{\rm s}^{\rm exp}$ for the singlet and the triplet states
(when there are reliable data for both). \\
Furthermore, the sensitivity of the effective $Q$'s determined
from the specific coupling (\ref{ARC1L}), has been
checked by analyzing the deviation of each $Q_a$ for a $25\%\,$ shift of
$\Lambda_{n_f=3}^{(1,{\rm eff})}$ around the value 193\ MeV, and one finds
that the average change in the momentum scale amounts to~$3\%\,$.
This makes the resulting
$\alpha_{\rm s}^{\rm exp}(Q^2)$ reliable, at least qualitatively, even in
the deep IR region ($Q<0.2\,$GeV), where the discrepancy with respect
to massless $\alpha_{\rm E}^{(1)}(Q^2)\,$ is sizable.\\
There is a subtle point concerning the choice of
the ``unperturbed'' $\alpha_{\rm s}$ involved in the static
Hamiltonian (\ref{eq:static}). Actually, the value adopted is very near
to the $\alpha_{a}^{\rm th}$ pertaining to the $b\bar{b}(1S)$ state,
but definitively smaller than the typical
$\alpha_{a}^{\rm  th}\,$'s. 
The point is that the hyperfine splitting is much
more sensible than the c.o.g.\ mass to the behaviour of the unperturbed
wave function at small distance (large momentum), which is specifically
controlled by the value of the unperturbed~$\alpha_{\rm s}\,$. As a result,
the effective fixed value $ \alpha^{\rm spl}_{\rm s} $ in
Eq.\ (\ref{eq:static}) that
reproduces the same splitting as by using the coupling
$ \alpha_{\rm E}(Q^2) $
turns out to
be significantly smaller than $\alpha_{a}^{\rm th}$ calculated from
the c.o.g.\ mass. Essentially, it was chosen a phenomenological value
for the unperturbed $\alpha_{\rm s}$ in order to have a good reproduction
of the hyperfine splitting so as to reasonably reconstruct the
c.o.g.\ of the doublet when one component is missing. It was then used
the position of the c.o.g.\ (which is rather stable w.r.t.\
the unperturbed $\alpha_{\rm s}$) to extract
$\alpha_{\rm s}^{\rm exp}(Q_a^2)\,$.\\
We finally stress again that the 1-loop analytic coupling with the above
mentioned value of the scaling constant used in our computation, Eq.\
(\ref{ARC1L}), differs by no more than $1\%$ from
the 3-loop analytic coupling in the region $0.5<Q<1.2\,$GeV where all
the input states ($\pi$, $\phi\,$, $J/\psi\,$ and $\Upsilon\,$) fall.


\section{BS-model results: concert of low and high energy data via APT}

All results are displayed in details in tables I-VII of App.~B, and
pictorially in Fig.\ \ref{low}
taken from Ref.~\cite{Baldicchi:2007ic}.
The first three
columns specify the
state and its experimental mass as given by~\cite{pdg}. The fourth
column gives our theoretical results for the  meson masses, and the
last three give the effective $Q\,$, the relative 3-loop APT coupling
$\alpha_{\rm E}^{(3)}(Q^2)$ and the 
experimental coupling with errors (theoretical and experimental).\\
In Fig.\ \ref{low} values of $\alpha_{\rm s}^{\rm exp}$ at the same Q
from triplet and singlet states have been combined through a weighted
average according to their errors (both experimental and theoretical).
The c.o.g.\ of the light-heavy states (that are interpreted according to
the j-j scheme) are also reported in the above figure.
As can be seen, the agreement between the 3-loop analytic coupling
with $\Lambda_{n_f=3}^{(3)}=417\,$MeV, and the points representing our
experimental values for the QCD coupling is quite good
within the errors down to 200~MeV.\\
At energies below 200~MeV a tendency
of $\alpha_{\rm s}^{\rm exp}(Q^2)$ to diminish with $Q$ seems to exist.
As already noted, such a deep IR behaviour could be
theoretically understood within theoretical models, in particular
within the ``massive'' modification of APT in Sec.~2.
Specifically, as displayed in Fig.~\ref{low}, the
one-loop coupling $\alpha(Q^2, m^2)$~(\ref{MARC}) with an
effective mass $m\ind{}{eff} \simeq (38 \pm 10)\,$MeV reasonably fits
all experimental points down to the very low Q region.\\
Let us notice, however, that the analysis of such an extreme IR behaviour
is based on high orbital excitations (D and F states),
lying well above the strong decay thresholds and with large widths.
As a consequence, the theoretical reliability of the method is lower
at these scales, as apparent from the large estimated errors.
Moreover, also the discrepancy between $\alpha_{\rm E}^{(1)}(Q^2)$ (used
in the calculation) and $\alpha_{\rm E}^{(3)}(Q^2)$ (used as a reference
term) rises above 10$\%$ at these scales. In fact, only the two states
$\pi_2$(1670) (interpreted as $s\bar s\,(1^1D_2)$) and $f_2$(2150)
($s\bar s\,(1^3F_2)$), corresponding to $Q\sim120\,$MeV, generate
$\alpha_{\rm s}^{\rm exp}(Q^2)$ (marginally) out of the error bands, and
the state $f_2$(2150) (observed only once) has never been confirmed.\\
Restricting our considerations to a sample of better established data,
which excludes high orbital excitations as D and F states, the comparison
with the BS meson masses yields a $\chi^2\sim1$ if an additional conventional 
error of 20~MeV is assigned to the latters. This error should account
for the sources of theoretical uncertainty not explicitly evaluated, and
produces an average additional error of roughly $5\%$ on 
$\alpha_{\rm s}^{\rm exp}(Q^2)\,$.\\
At this point it is worthwhile to comment on the dependence of the
results on the renormalization scheme. First of all, our definition
of the coupling is implicitly contained in ansatz~(\ref{eq:wilson}).
Specifically, here one assumes both that $i\ln W$ is dominated by the OGE term
after the subtraction of the area term, and that the OGE term is
represented as the corresponding tree-level expression, the fixed
coupling~$\al{}{s}$ being replaced with the running
one~$\al{}{s}(Q^2)$. The latter assumption amounts to the embodying
all the dressing effects into $\al{}{s}(Q^2)$ (see, e.g., Sec.~3.2 of
Ref.~\cite{Prosperi:2006hx} and references cited therein). It is worth
noting that the coupling defined in such way is free of unphysical
singularities by
construction. At the same time, the analytic running coupling
$\al{}{E}(Q^2)$, which is involved in our calculations, is remarkably
stable with respect to both the higher loop corrections and the
choice of renormalization scheme (see Sec.~2 and detailed
discussion of this issue in Ref.~\cite{APT-07}). Thus one might expect that
the same situation should also occur for~$\al{}{exp}(Q^2)$, with the
possible exception for the deep infrared region (see, e.g., Sec.~4.5
of Ref.~\cite{Prosperi:2006hx} and references cited therein), where other
nonperturbative effects could be relevant.\\
Notice that in our selection of states as a rule we have excluded irregular
and incomplete multiplets. Of this type, e.g., in the light quark sector,
are the $3S$ states ($ m_{3\,^3S_1}-m_{3\,^1S_0} $
is anomalously large and about twice as
$ m_{2\,^3S_1}-m_{2\,^1S_0} $), $1\,^3P$ ($ m_{1\,^3P_0} $ being larger than
$ m_{1\,^3P_1} $), $1\,^3D$, $F$, $G$, $H$ (incomplete).
If however included in the
analysis, all these states would bring the results in agreement with the
general tendency outlined.\\
Finally, Fig.~\ref{tot}
(taken from Ref.~\cite{Baldicchi:2007ic})
displays a synthesis of results for
$\alpha_{\rm s}(Q^2)$ defined from bound states in the BS framework
with high energy data. Here, low energy results are reported in a
logarithmic scale from 100~MeV to 220~GeV together with a sample of
high energy data as given by S.~Bethke \cite{Bethke:2006ac}, against
the 3-loop analytic coupling $\alpha^{(3)}_{\rm E}(Q^2)$~(\ref{ARC})
and its massive modification~(\ref{MARC}). Also shown in the figure
is the common perturbative 3-loop coupling with IR singular behaviour
that is definitively ruled out by the data. As can be seen, the
BS-APT theoretical scheme allows a rather satisfactory correlated
understanding of very high and rather low energy phenomena.


\section{Conclusive remarks}
To summarize, we have exploited calculations within the
 Bethe-Salpeter formalism adjusted for QCD, in order to extract an
 ``experimental'' strong coupling
$\alpha_{\rm s}^{\rm exp}(Q^2)$ below 1~GeV
 by comparison with the meson spectrum.\\
This work extends the analysis given in~\cite{Baldicchi:2007ic},
providing technical details, the complete set of numerical results
and their thorough discussion.\\
  A key point is the comparison of $\alpha_{\rm s}^{\rm exp}$
with the analytic coupling
$ \alpha_{\rm E}(Q^2) $
which avoids the hurdle of
the unphysical singularities in the IR region~\cite{ShSol96-7}.\\
The method consists in solving the eigenvalue equation for the
squared mass operator as given by Eq.~(\ref{eq:M2}), obtained by a
three dimensional reduction of the original BS equation. The
relativistic potential $U$ then follows from a proper
ansatz~(\ref{eq:wilson}) on the Wilson loop to encode confinement,
and is the sum~(\ref{eq:M2}) of a one-gluon-exchange term $U_{\rm
OGE}$ and a confining term $U_{\rm Conf}\,$. The coupling occurring
in the perturbative part of the potential needs to be IR finite since
its argument has to be identified with the momentum transfer in the
$q\bar{q}$ interaction, and this typically takes values down to few
hundreds MeV. The usual perturbative running coupling
$\bar{\alpha}_{\rm s}$ has then been replaced by the 1-loop analytic
expression $\alpha_{\rm E}^{(1)}$ Eq.~(\ref{ARC1L}) with an
effective QCD scale $\Lambda_{n_f=3}^{(1, \rm{eff})}=193$~MeV. This
value reasonably reproduces the 3-loop analytic coupling, normalized at
the Z~boson mass along with world average~\cite{pdg} (i.e.,
$\Lambda_{n_f=5}^{(3)}=236\,$MeV which leads to
$\Lambda_{n_f=3}^{(3)}=417\,$MeV by continuous threshold matching, see App.~C).\\
Thus we have taken advantage of our BS results for the meson
spectrum, both in the light and heavy quark sector, to infer within
this framework the fixed coupling value for each state that exactly
matches the theoretical and experimental mass.\\
Our results are twofold. On the one hand, as expected (given the good
agreement of theoretical and experimental meson data), the 3-loop
analytic coupling reasonably fits $\alpha_{\rm s}^{\rm exp}(Q^2)$
from 1~GeV down to 200~MeV within the estimated theoretical and
experimental errors, with very few exceptions. This confirms and
yields a quantitative estimate of the relevance of the APT
to IR phenomena down to 200~MeV.\\
On the other hand, below this scale, the experimental points
 exhibit a tendency to fall under the APT curve.
This could give a hint either 
 on the vanishing of $\alpha_{\rm s}(Q^2)$ as $Q\to0\,$ (in
 concert with some results from lattice simulations~\cite{Lattice}), 
or on the existence of a finite IR limit of $\alpha_{\rm s}(Q^2)$ 
lower than the universal APT freezing value.
The former IR behaviour can be theoretically understood in
the framework of a recent ``massive'' modification \cite{MAPT} of
 the APT algorithm, which takes into account effects of a finite
 threshold in the dispersion relation. Since in the extremely low
 $Q$ region confinement forces play the dominant role, the
 reasonable agreement between the ``massive'' APT model and the results
 of the BS formalism would suggest a relation between the linear
 potential, arising from the area term in the ansatz (\ref{eq:wilson}),
and the thresholds effects in the analytic
 properties of the QCD coupling, to be further investigated.

\section*{Acknowledgements}

The partial support of grants RFBR 05-01-00992, NS-5362.2006.2, and
BelRFBR F06D-002 is acknowledged. We would like to thank Professor D.~V.~Shirkov
for useful discussions.

\appendix


\section{Second order Bethe-Salpeter formalism}

In the QCD framework a {\it second order} four point quark-antiquark
function and full quark propagator can be defined as
\begin{equation}
H_{4}(x_1,x_2;y_1,y_2)= -{1\over 3} {\rm Tr _{color}}
\langle \Delta_1 (x_1,y_1;A)
{\Delta}_2(y_2,x_2;A)\rangle
\label{eq:so4point}
\end{equation}
\noindent
and
\begin{equation}
H_{2}(x-y) = {i \over \sqrt{3}}{\rm Tr_{color}}
\langle \Delta(x,y;A)\rangle \,,
\label{eq:so2point}
\end{equation}
where
\begin{equation}
\langle f[A] \rangle = \int  DA\, M_F [A]\, e^{iS_G[A]} f[A]  \,,
\label{eq:expt}
\end{equation}
\noindent
$
M_F[A] = {\rm Det} \, \Pi_{j=1}^2 [1 + g\gamma^\mu A_\mu
( i\gamma_j^\nu \partial_{j\nu} - m_j^{\rm curr})^{-1}]
$
and $\Delta (x,y;A)$ is the {\it second order} quark
propagator in an external gauge field.

The quantity $ \Delta $ is defined by
the second order differential equation
\begin{equation}
(D_\mu D^\mu +m^2_{\rm curr} -{1\over 2} g \, \sigma^{\mu \nu} F_{\mu \nu})
\Delta (x,y;A) = -\delta^4(x-y) \, ,
\label{eq:soprop}
\end{equation}
($\sigma^{\mu \nu} = {i\over 2} [\gamma^\mu, \gamma^\nu]$ and
$D_\mu=\partial_\mu + ig A_\mu$) and it is related to
the corresponding first order propagator by
$
S(x,y;A) = (i \gamma^\nu D_\nu + m_{\rm curr}) \Delta (x,y;A) \,
$, $m_{\rm curr}$ being the so-called current mass of the quark.

The advantage of considering second order quantities is that the spin
terms are more clearly separated and it is possible to write for
$\Delta$ a generalized Feynman-Schwinger representation, i.e., to
solve Eq.~(\ref{eq:soprop}) in terms of a quark path integral
\cite{BMP,BP}. Using the latter in (\ref{eq:so4point}) or
(\ref{eq:so2point}) a similar representation can be obtained for
$H_{4}$ and $H_{2}$.

The interesting aspect of this final representation is that the gauge field
appears in it only through a Wilson line correlator $W$.
In the limit $x_2 \to x_1$, $y_2 \to y_1$ or $y \to x$ the Wilson lines
close in a single Wilson loop $\Gamma$
and if $\Gamma$ stays on a plane,
$i\ln W$ can be written according to (\ref{eq:wilson}) as

\begin{eqnarray}
&& i\ln W = {16\pi\over 3}\alpha_{\rm s} \oint dz^\mu \oint dz^{\nu \prime}
D_{\mu \nu}(z-z^\prime) +
\label{eq:wils} \\
&& \sigma \oint dz^0 \oint dz^{0 \prime} \delta (z^0-z^{0\prime})
|{\bf z} - {\bf z}^\prime| \int_0^1 d\lambda
 \Big \{ 1 -  [\lambda {d{\bf z}_\perp \over dz^0}
 + (1-\lambda) {d{\bf z}_\perp ^\prime \over dz^{0 \prime}} ]^2
\Big \}^{1\over 2} \, . \nonumber
\end{eqnarray}
The area term here is written as the algebraic sum of successive
equal time strips and $ d{\bf z}_\perp = d{\bf z} -
(d{\bf z}\cdot {\bf r}){\bf r}/r^2 $ denotes the transversal component of
$ d{\bf z} $. The basic assumption now
is that in the center of mass frame
(\ref{eq:wils}) remains a good approximation even
in the general case, i.e.,
for non flat curves and  when $x_2 \ne x_1$,
$y_2 \ne y_1$ or $y \ne x$.
Then, by appropriate manipulations on the resulting expressions,
an inhomogeneous
Bethe-Salpeter equation for the 4-point function
$H_{4}(x_1,x_2;y_1,y_2)$ and a Dyson-Schwinger equation for
$H_{2}(x-y)$ can be derived in a kind of generalized ladder and rainbow
approximation. This should appear plausible, even from the point of
view of graph resummation, for the analogy between the perturbative
and the confinement terms in (\ref{eq:wils}).

In momentum representation, the corresponding homogeneous
BS-equation becomes
\begin{eqnarray}
\Phi_P (k) &=& -i \int {d^4u \over (2 \pi)^4} \;
   \hat I_{ab} \left( k-u; \, {1 \over 2}P
   +{k+u \over 2}, \,
   {1 \over 2}P-{k+u \over 2} \right)\,\,\times \nonumber \\
    & & \qquad\times\,\,
   \hat H_{2}^{(1)}   \left({1 \over 2} P  + k \right)
      \sigma^a  \, \Phi_P (u) \, \sigma^b \,
   \hat H_{2}^{(2)} \left(-{1 \over 2} P + k \right) \, ,
\label{eq:bshom}
\end{eqnarray}
\noindent
where $\sigma^0=1$; $a, \, b = 0, \, \mu\nu$;
the c.m.\ frame has to be understood, $P=(m_B, {\bf 0})$;
$\Phi_P (k)$ denotes the appropriate {\it second order} wave function,
that in terms of the second order field $\phi (x) = (i\gamma^\mu
D_\mu + m_{\rm curr})^{-1}\psi(x)$ can be defined as the Fourier
transform of
$
  \langle 0|\phi({\xi \over 2}) \bar\psi(-{\xi \over 2}) |P\rangle\,
$.

Similarly, in terms of the irreducible self-energy, defined by
$\hat H_{2}(k) = i (k^2-m_{\rm curr}^2)^{-1} +
i ( k^2-m_{\rm curr}^2)^{-1} \, i \,
\Gamma (k) \, \hat H_{2}(k) \,$,
the Dyson-Schwinger equation can be written
\begin{equation}
\hat \Gamma(k) =  \int {d^4 l \over (2 \pi)^4}  \,
\hat I_{ab} \Big ( k-l;{k+l \over 2},{k+l \over 2} \Big )
\sigma^a \hat H_{2}(l) \, \sigma^b \ .
\label{eq:dshom}
\end{equation}

The kernels are the same in the two Eqs.\
(\ref{eq:bshom}) and (\ref{eq:dshom}), consistently with the
requirement of chiral symmetry limit \cite{chiral}, being given by
\begin{eqnarray}
& & \hat I_{0;0} (Q; p, p^\prime)  =
   16 \pi {4 \over 3} \alpha_{\rm s} p^\alpha p^{\prime \beta}
  \hat D_{\alpha \beta} (Q)  + \nonumber \\
& &  \quad + 4 \sigma  \int \! d^3 {\bf \zeta} e^{-i{\bf Q}
   \cdot {\bf \zeta}}
    \vert {\bf \zeta} \vert \epsilon (p_0) \epsilon ( p_0^\prime )
   \int_0^1 \! d \lambda \{ p_0^2 p_0^{\prime 2} -
   [\lambda p_0^\prime {\bf p}_{\rm T} +
   (1-\lambda) p_0 {\bf p}_{\rm T}^\prime ]^2 \} ^{1 \over 2} \nonumber \\
& & \hat I_{\mu \nu ; 0}(Q;p,p^\prime) = 4\pi i {4 \over 3} \alpha_{\rm s}
   (\delta_\mu^\alpha Q_\nu - \delta_\nu^\alpha Q_\mu) p_\beta^\prime
   \hat D_{\alpha \beta}(Q)  - \nonumber \\
& & \qquad \qquad \qquad  - \sigma  \int d^3 {\bf \zeta} \, e^{-i {\bf Q}
\cdot {\bf \zeta}} \epsilon (p_0)
   {\zeta_\mu p_\nu -\zeta_\nu p_\mu \over
   \vert {\bf \zeta} \vert \sqrt{p_0^2-{\bf p}_{\rm T}^2}}
   p_0^\prime  \nonumber \\
& & \hat I_{0; \rho \sigma}(Q;p,p^\prime) =
   -4 \pi i{4 \over 3} \alpha_{\rm s}
   p^\alpha (\delta_\rho^\beta Q_\sigma - \delta_\sigma^\beta Q_\rho)
   \hat D_{\alpha \beta}(Q) + \nonumber \\
& & \qquad \qquad  \qquad  + \sigma  \int d^3 {\bf \zeta} \, e^{-i{\bf Q}
  \cdot {\bf \zeta}} p_0
  {\zeta_\rho p_\sigma^\prime - \zeta_\sigma p_\rho^\prime \over
  \vert {\bf \zeta} \vert \sqrt{p_0^{\prime 2}
   -{\bf p}_{\rm T}^{\prime 2}} }
  \epsilon (p_0^\prime)  \nonumber \\
& & \hat I_{\mu \nu ; \rho \sigma}(Q;p,p^\prime) =
   \pi {4\over 3} \alpha_{\rm s}
  (\delta_\mu^\alpha Q_\nu - \delta_\nu^\alpha Q_\mu)
  (\delta_\rho^\alpha Q_\sigma - \delta_\sigma^\alpha Q_\rho)
  \hat D_{\alpha \beta}(Q) \, ,
\label{eq:imom}
\end{eqnarray}
\noindent
where in the second and in the third equation $\zeta_0 = 0$ has to be
understood. Notice that, due to the privileged role given to the c.m.\ frame,
the terms proportional to $\sigma$ in (\ref{eq:imom}) formally are not
covariant.\\
In fact, it can be checked that $\Gamma(k)$ can
be consistently assumed to be spin independent and
Eq.~(\ref{eq:dshom}) can be rewritten in the simpler form
\begin{equation}
\Gamma (k) = i \int {d^4l \over (2 \pi)^4} \,
  {R(k,l) \over l^2-m^2+ \Gamma(l)} ,
\label{eq:dssc}
\end{equation}
\noindent
with
\begin{eqnarray}
R(k,l)&=& 4\pi {4 \over 3} \alpha_{\rm s}
\left [(k+l)^{\mu}(k+l)^{\nu}D_{\mu\nu}(k-l)+\right.
\nonumber\\
& &\left.(k-l)_{\nu}(k-l)^{\nu}D_{\mu}^{\,\,\mu}(k-l)
-(k-l)^{\mu}(k-l)^{\nu}D_{\mu\nu}(k-l)\right]
+\nonumber \\
  & & + \sigma \int d^3 {\bf r} e^{-i({\bf k}-{\bf l})
      \cdot{\bf r}} r (k_0 + l_0)^2 \sqrt{1-{({\bf k_\perp}+{\bf
      l_\perp})^2 \over (k_0 + l_0)^2}} \,,
\label{eq:dskernel}
\end{eqnarray}
\noindent
${\bf k_\perp}$ and ${\bf l_\perp}$ denoting as above the transversal
part of ${\bf k}$ and ${\bf l}$. Eq.\ (\ref{eq:dssc}) can be solved
by iteration resulting in an expression of the form $\Gamma(k^2,\bf{k}^2)$,
since~(\ref{eq:dskernel}) is not formally covariant.  Then
the constituent (pole) mass $m$ is defined by the
equation
\be
m^2 -m^2_{\rm curr}+\Gamma(m^2,{\bf k}^2)=0
\lb{mcurr}
\ee
and the dependence on ${\bf k}^2\,$,
being an artifact of the ansatz~(\ref{eq:wilson}),
is eliminated by extremizing $m(\bf{k}^2)$ in $\bf{k}^2\,$.

The 3-dimensional reduction of Eq.~(\ref{eq:bshom}) can be
obtained by a usual procedure of replacing $H_2(k)$ with
$i (k^2-m^2)^{-1} $
and $\hat{I}_{ab}$ with its so-called instantaneous approximation
$ \hat{I}_{ab}^{\rm inst}({\bf k}, {\bf u})\,$.
In this way, one can explicitly integrate over $u_0$ and arrive to
a 3-dimensional equation in the form of the eigenvalue equation
for a squared mass operator Eq.~(\ref{eq:M2}), with \cite{BMP}
\begin{equation}
   \langle {\bf k} \vert U \vert {\bf k}^\prime \rangle =
        {1\over (2 \pi)^3 }
        \sqrt{ w_1 + w_2 \over 2  w_1  w_2} \; \hat I_{ab}^{\rm \; inst}
        ({\bf k} , {\bf k}^\prime) \; \sqrt{ w_1^\prime + w_2^\prime \over 2
         w_1^\prime w_2^\prime}\; \sigma_1^a \sigma_2^b \,.
\label{eq:quadrrel}
\end{equation}
Finally by using Eq.~(\ref{eq:imom}) one obtains
Eqs.~(\ref{upt}-\ref{eq:uconf1}).

  Alternatively, in more usual terms, one could look for the eigenvalue of the
mass operator or center of mass Hamiltonian
$          H_{\rm CM} \equiv M = M_0 + V  $
with $V$ defined by $M_0V+VM_0+V^2=U$. Neglecting term $V^2$
the linear form potential $V$ can be
obtained from $U$ by the replacement
$
\sqrt{ (w_1+w_2) (w_1^\prime +w_2^\prime)\over w_1w_2w_1^\prime w_2^\prime}
\to {1\over 2\sqrt{w_1 w_2 w_1^\prime w_2^\prime}}
$.
The resulting
expression is particularly useful for a comparison with models based
on potential. In particular, in the static limit $V$ reduces to the Cornell
potential
\begin{equation}
V_{\rm stat} = - {4 \over 3} {\alpha_{\rm s} \over r } + \sigma r \, .
    \label{eq:static2}
\end{equation}
Note that it is necessary to introduce a cut-off $B$ in
Eq.~(\ref{eq:dssc}). As a consequence the constituent mass turns out
to be a function of the current mass and of $B$, $m=m\,(m_{\rm curr},
B)$. Then if one uses a running current mass $m_{\rm curr}(Q^2)$ we
obtain a running constituent mass $m(Q^2)$ as it has been done
in~\cite{Baldicchi:2004wj} (see also Ref.~\cite{PRD}). However the
singular expression used there
\begin{equation}
 m_{\rm curr}(Q^2) = {\hat m} \left( \frac{Q^{2}/ \Lambda^{2} -1}
{Q^2/ \Lambda^2 \ln ( Q^2/ \Lambda^2 )} \right)^{ \gamma_0/
2 \beta_0}
\end{equation}
is not consistent with Eq.~(\ref{ARC1L}),
and if a more consistent assumption is taken, e.g.,
\begin{equation}
 m_{\rm curr}(Q^2) = {\hat m} \left(
\alpha_{\rm E}^{(1)} (Q^2) \right)^{ \gamma_0/
2 \beta_0}\,,
\end{equation}
or the other resulting from the analytization of a similar expression
with $\alpha_{\rm E}^{(1)}(Q^2)$ replaced with the ordinary
perturbative $\alpha_{\rm s}^{(1)}(Q^2)$, the dependence of $m$ on
$Q^2$ is strongly reduced. For this reason even the light quark mass
is here treated as a constants to be adjusted with the the
data\footnote{In this way the only role that is left to the DS
equation is to justify the difference between the constituent and the
current masses.}.


\newpage
\section{Numerical results}

The tables below display the complete set of results as explained in
Sec.~5. We recall the values of all the parameters:
$\sigma=0.18\;{\rm GeV}^{2}\,$, $\Lambda_{n_f=3}^{(1,{\rm
eff})}=193\;{\rm MeV}\,$, $m_{q}=196\;{\rm MeV}\;(q=u,d)\,$,
$m_{s}=352\;{\rm MeV}\,$, $m_{c}=1.516\;{\rm GeV}\,$ and
$m_{b}=4.854\; {\rm GeV}\,$. Meson masses are given in MeV. The last
column displays the experimental coupling $\alpha_{\rm s}^{\rm
exp}(Q_a^2)$ with the theoretical error $\Delta_{\rm NLO}$ due to 
the next-to-leading order terms neglected, the
theoretical error $\Delta_{\rm \Gamma}$ from the half width $
\Gamma/2 $ and the experimental error $\Delta_{\rm exp}$
respectively.\\
\\
{\footnotesize \dag~Center of gravity masses of the incomplete
multiplets estimated in analogy with other multiplets.}

\begin{table} [!ht]
\footnotesize{
\caption{$q \bar{q} \;\, (q = u,d) $
\vspace{0.7cm} } \hspace{-2.truecm}\begin{tabular}{ccccccc}
  &  &  &  &  &  & \\
States & (MeV) & $m_{\rm exp}$ & $m_{\rm th}$ & $Q$ &
$\alpha_{\rm E}^{(3)}$ &
$\alpha_{\rm s}^{\rm exp}\pm\Delta_{\rm NLO}\pm\Delta_{\rm \Gamma}\pm
 \Delta_{\rm exp}$
\\
\hline
$ 1 \, {^{1} {\rm S}_{0}} $ &
$
\left\{
\begin{array}{c}
\pi^{0}   \\
\pi^{\pm}
\end{array}
\right.
$
&
$
\left.
\begin{array}{c}
134.9766 \pm 0.0006  \\
139.57018 \pm 0.00035
\end{array}
\right\}
138
$ & 136 & 401 & 0.522 & $ 0.534 \pm 0.122 \pm - \pm - $  \\
$ 1 \, {^{3} {\rm S}_{1}} $ & $ \rho (770) $ & 775.5 $ \pm $ 0.4 &
749 & &  & $ 0.517 \pm 0.122 \pm 0.048 \pm - $ \\
$ 1 \Delta {\rm SS} $ &  & 638 & 613 &   &  &  \\
$ 2 \, {^{1} {\rm S}_{0}} $ & $ \pi (1300) $ & 1300 $ \pm $ 100 &
1223 & 448 & 0.502 & $ 0.451 \pm 0.114 \pm 0.152 \pm 0.081 $  \\
$ 2 \, {^{3} {\rm S}_{1}} $ & $ \rho (1450) $ & 1459 $ \pm $ 11 &
1363 &  &  & $ 0.427 \pm 0.114 \pm 0.062 \pm 0.010 $  \\
$ 2 \Delta {\rm SS} $ &  & 159 & 139 &  &  &   \\
\hline
$ 1 \, {^{1} {\rm P}_{1}} $ & $ b_{1} (1235) $
& 1229.5 $ \pm $ 3.2 & 1234 & 209 & 0.637 & $ 0.688 \pm 0.155
\pm 0.124 \pm 0.006 $ \\
\hline
$ 1 \, {^{1} {\rm D}_{2}} $ & $ \pi_{2} (1670) $ & $ 1672.4
\pm 3.2 $ & 1595 & 144 &
 0.701 & $ 0.544 \pm 0.151 \pm 0.364 \pm 0.009 $ \\
\end{tabular}
\vspace{-0.2truecm}}
\end{table}

\begin{table}[!ht] 
\footnotesize{
\caption{ $ s \bar{s} $
\vspace{3.5cm}}
\hspace{-2.truecm}\begin{tabular}{ccccccc}
  &  &  &  &  &  &  \\
States & (MeV) & $m_{\rm exp}$ & $m_{\rm th}$ & $Q$ & $
\alpha_{\rm E}^{(3)}$ &
$\alpha_{\rm s}^{\rm exp}\pm\Delta_{\rm NLO}\pm\Delta_{\rm \Gamma}\pm
 \Delta_{\rm exp}$
\\
\hline
$ 1 \, {^{3} {\rm S}_{1}} $ & $ \phi (1020) $
& 1019.460 $ \pm $ 0.019 & 1019 & 418 & 0.514 &
$ 0.525 \pm 0.098 \pm 0.002 \pm - $  \\
$ 2 \, {^{3} {\rm S}_{1}} $ & $ \phi (1680) $ & 1680 $ \pm $ 20
& 1602 & 454 & 0.500 & $ 0.435 \pm 0.096 \pm 0.068 \pm 0.019 $ \\
\hline
$ 1 \, {^{1} {\rm P}_{1}} $ & $ h_{1} (1380) $
& 1386 $ \pm $ 19 & 1472 & 216 & 0.631 & $ 0.824 \pm 0.098
\pm 0.083 \pm 0.032 $ \\
$
\begin{array}{c}
1 \, {^{3} {\rm P}_{2}} \\
1 \, {^{3} {\rm P}_{1}} \\
1 \, {^{3} {\rm P}_{0}}
\end{array}
$ &
$
\begin{array}{c}
f_{2}^{\prime} (1525) \\
f_{1} (1510) \\
f_{0} (1500)
\end{array}
$ &
$
\left.
\begin{array}{c}
1525 \pm 5 \\
1518 \pm 5 \\
1507 \pm 5
\end{array}
\right\}
1521
$ & 1484 &  &  & $ 0.603 \pm 0.098 \pm 0.070 \pm 0.009 $  \\
\hline
$ 1 \, {^{1} {\rm D}_{2}} $ & $ \eta_{2} (1870) $ &
$ 1842 \pm 8 $ & 1807 & 149 & 0.695 &
$ 0.658 \pm 0.079 \pm 0.318 \pm 0.023 $ \\
\hline
$
\begin{array}{c}
1 \, {^{3} {\rm F}_{4}} \\
1 \, {^{3} {\rm F}_{3}} \\
1 \, {^{3} {\rm F}_{2}}
\end{array}
$ &
$
\begin{array}{c}
  \\  \\ f_{2} (2150)  \\
\end{array}
$ &
$
\left.
\begin{array}{c}
    \\   \\ 2156 \pm 11  \\
\end{array}
\right\} 2165^{\dag} $ &
2070 & 118 & 0.733 & $ 0.452 \pm 0.064 \pm 0.137 \pm 0.024 $ \\
\end{tabular}
}
\end{table}

\begin{table}[!ht] 
\footnotesize{
\caption{ $ q \bar{s} \; (q = u,d) $
\vspace{0.5cm}} \hspace{-2.truecm}\begin{tabular}{ccccccc}
& & & & & & \\
States & (MeV) & $m_{\rm exp}$ & $m_{\rm th}$ & $ Q $ & $
\alpha_{\rm E}^{(3)} $ &
$ \alpha_{\rm s}^{\rm exp}\pm\Delta_{\rm NLO}\pm\Delta_{\rm \Gamma}\pm
 \Delta_{\rm exp} $
\\
\hline
$ 1 \, {^{1} {\rm S}_{0}} $ &
$
\left\{
\begin{array}{c}
K^{0}   \\
K^{\pm}
\end{array}
\right.
$
&
$
\left.
\begin{array}{c}
497.648 \pm 0.022  \\
493.677 \pm 0.016
\end{array}
\right\}
495
$ & 491 & 409 & 0.518 & $ 0.529 \pm 0.122 \pm - \pm - $  \\
$ 1 \, {^{3} {\rm S}_{1}} $ &
$
\left\{
\begin{array}{c}
 K^{\ast} (892)^{0}   \\
 K^{\ast} (892)^{\pm}
\end{array}
\right.
$
&
$
\left.
\begin{array}{c}
 896.00 \pm 0.25 \\
 891.66 \pm 0.26
\end{array}
\right\}
893.11 $ & 887 &  &  & $ 0.526 \pm 0.122 \pm 0.017 \pm - $ \\
$ 1 \Delta {\rm SS} $ &  & 398 & 396 &  &  &  \\
$ 2 \, {^{3} {\rm S}_{1}} $ & $ K^{\ast} (1410) $ & 1414 $ \pm $ 15 &
1485 & 451 & 0.501 & $ 0.571 \pm 0.117 \pm 0.102 \pm 0.013 $ \\
\hline
$ 1 \, {^{1} {\rm P}_{1}} $ & $ K_{1} (1270) $ & 1272 $ \pm $ 7
& 1355 & 213 & 0.634 & $ 0.820 \pm 0.129 \pm 0.081 \pm 0.012 $ \\
$
\begin{array}{c}
1 \, {^{3} {\rm P}_{2}}
\vspace{3mm}
 \\
1 \, {^{3} {\rm P}_{1}} \\
1 \, {^{3} {\rm P}_{0}}
\end{array}
$ &
$
\begin{array}{c}
\left\{
\begin{array}{c}
K_{2}^{\ast} (1430)^{0} \\ K_{2}^{\ast} (1430)^{\pm}
\end{array}
\right. \\
K_{1} (1400) \\
K_{0}^{\ast} (1430)
\end{array}
$ &
$
\left.
\begin{array}{c}
1432.4 \pm 1.3 \\
1425.6 \pm 1.5 \\
1402 \pm 7 \\
1414 \pm 6
\end{array}
\right\} 1417.7 $ & 1367 &  &  & $ 0.583 \pm 0.129 \pm 0.133 \pm 0.007 $ \\
\hline
$
\begin{array}{c}
1 \, {^{3} {\rm D}_{3}} \\
1 \, {^{3} {\rm D}_{2}} \\
1 \, {^{3} {\rm D}_{1}}
\end{array}
$ &
$
\begin{array}{c}
K_{3}^{\ast} (1780) \\
K_{2} (1770) \\
K^{\ast} (1680)
\end{array}
$ &
$
\left.
\begin{array}{c}
1776 \pm 7  \\
1773 \pm 8  \\
1717 \pm 27
\end{array}
\right\}
1763
$ & 1712 & 150 & 0.694 & $ 0.617 \pm 0.113 \pm 0.273 \pm 0.031 $ \\
\hline
$
\begin{array}{c}
1 \, {^{3} {\rm F}_{4}} \\
1 \, {^{3} {\rm F}_{3}} \\
1 \, {^{3} {\rm F}_{2}}
\end{array}
$ &
$
\begin{array}{c}
K_{4}^{\ast} (2045) \\
K_{3} (2320) \\
K_{2}^{\ast} (1980)
\end{array}
$ &
$
\left.
\begin{array}{c}
2045 \pm 9  \\
2324 \pm 24  \\
1973 \pm 25
\end{array}
\right\} 2121 $ &
1973 & 116 & 0.736 & $ 0.248 \pm 0.095 \pm 0.413 \pm 0.071 $ \\
\end{tabular}
}
\vspace{9.truecm}
\end{table}

\newpage

\begin{table}[!ht] 
\footnotesize{
\caption{ $ c \bar{c} $
\vspace{0.05cm}}
\hspace{-2.truecm}\begin{tabular}{ccccccc}
& & & & & & \\
States & (MeV) & $m_{\rm exp}$ & $m_{\rm th}$ &
$ Q $ & $ \alpha_{\rm E}^{(3)} $&
$ \alpha_{\rm s}^{\rm exp}\pm\Delta_{\rm NLO}\pm\Delta_{\rm \Gamma}\pm
 \Delta_{\rm exp} $ \\
\hline
$ 1 \, {^{1} {\rm S}_{0}} $ & $ \eta_{c} (1S) $
& 2980.4 $ \pm $ 1.2 & 2980 & 561 & 0.464 &
$ 0.467 \pm 0.025 \pm 0.008 \pm 0.001 $ \\
$ 1 \, {^{3} {\rm S}_{1}} $ & $ J/\psi (1S) $ & 3096.916 $ \pm $ 0.011
& 3097 &  &  & $ 0.467 \pm0.025 \pm - \pm - $ \\
$ 1 \Delta {\rm SS} $ &  & 117 & 118 &  &  &  \\
$ 2 \, {^{1} {\rm S}_{0}} $ & $ \eta_{c} (2S) $ & 3638 $ \pm $ 4
& 3595 & 500 & 0.483 & $ 0.446 \pm 0.023 \pm 0.007 \pm 0.004 $ \\
$ 2 \, {^{3} {\rm S}_{1}} $ & $ \psi (2S) $
& 3686.093 $ \pm $ 0.034 & 3653 &  &  & $ 0.455 \pm 0.023 \pm - \pm - $ \\
$ 2 \Delta {\rm SS} $ &  & 48 & 58 &  &  &   \\
$ 3 \, {^{3} {\rm S}_{1}} $ & $ \psi (4040) $ & 4039 $ \pm $ 1
& 4030 & 483 & 0.489 & $ 0.485 \pm 0.022 \pm 0.049 \pm 0.001 $ \\
$ 4 \, {^{3} {\rm S}_{1}} $ & $ \psi (4415) $ & 4421 $ \pm $ 4 &
4337 & 474 & 0.492 & $ 0.384 \pm 0.022 \pm 0.042 \pm 0.006 $ \\
\hline
$ 1 \, {^{1} {\rm P}_{1}} $ & $ h_{c} (1P) $ & 3525.93 $ \pm $ 0.27 &
3532 & 269 & 0.592 & $ 0.631 \pm 0.012 \pm - \pm - $  \\
$
\begin{array}{c}
1 \, {^{3} {\rm P}_{2}} \\
1 \, {^{3} {\rm P}_{1}} \\
1 \, {^{3} {\rm P}_{0}}
\end{array}
$ &
$
\begin{array}{c}
\chi_{c2} (1P) \\
\chi_{c1} (1P) \\
\chi_{c0} (1P)
\end{array}
$ &
$
\left.
\begin{array}{c}
3556.20 \pm 0.09 \\
3510.66 \pm 0.07 \\
3414.76 \pm 0.35
\end{array}
\right\} 3525.3 $ & 3537 &  &  & $ 0.640 \pm 0.012 \pm 0.002 \pm - $ \\
$
\begin{array}{c}
2 \, {^{3} {\rm P}_{2}} \\
2 \, {^{3} {\rm P}_{1}} \\
2 \, {^{3} {\rm P}_{0}}
\end{array}
$ &
$
\begin{array}{c}
\chi_{c2} (2P) \\
               \\
X(3872)
\end{array}
$ &
$
\left.
\begin{array}{c}
3929 \pm 5   \\
             \\
3871.2 \pm 0.5
\end{array}
\right\} 3915^{\dag} $ & 3929 & 274 & 0.589 &
$ 0.644 \pm 0.013 \pm 0.027 \pm 0.009 $ \\
\hline
$
\begin{array}{c}
1 \, {^{3} {\rm D}_{3}} \\
1 \, {^{3} {\rm D}_{2}} \\
1 \, {^{3} {\rm D}_{1}}
\end{array}
$ &
$
\begin{array}{c}
   \\
   \\
 \psi (3770)
\end{array}
$ &
$
\left.
\begin{array}{c}
   \\
   \\
3771.1 \pm 2.4
\end{array}
\right\} 3820^{\dag} $ &
3822 & 190 & 0.654 & $ 0.707 \pm 0.008 \pm 0.030 \pm 0.006 $ \\
$
\begin{array}{c}
2 \, {^{3} {\rm D}_{3}} \\
2 \, {^{3} {\rm D}_{2}} \\
2 \, {^{3} {\rm D}_{1}}
\end{array}
$ &
$
\begin{array}{c}
   \\    \\   \psi (4160)
\end{array}
$ &
$
\left.
\begin{array}{c}
   \\    \\
4153 \pm 3
\end{array}
\right\} 4183^{\dag} $ &
4150 & 198 & 0.646 & $ 0.606 \pm 0.009 \pm 0.132 \pm 0.008 $ \\
\end{tabular}
Quantum numbers of $ h_{c} (1P) $ and $ X(3872) $ mesons
are not well established.
}
\end{table}

\begin{table}[!ht] 
\vspace{-0.08truecm}
\footnotesize{
\caption{ $ b \bar{b} $
\vspace{0.07cm}}
\hspace{-2.6truecm}\begin{tabular}{ccccccc}
  & & & & & &  \\
States & (MeV) & $m_{\rm exp}$ & $m_{\rm th}$ & $ Q $ & $
\alpha_{\rm E}^{(3)} $ &
$ \alpha_{\rm s}^{\rm exp}\pm\Delta_{\rm NLO}\pm\Delta_{\rm \Gamma} \pm
 \Delta_{\rm exp} $
\\
\hline
$ 1 \, {^{3} {\rm S}_{1}} $ & $ \Upsilon (1S) $
& 9460.30 $ \pm $ 0.26 & 9461 & 951 & 0.381
& $ 0.378 \pm 0.006 \pm - \pm - $ \\
$ 2 \, {^{3} {\rm S}_{1}} $ & $ \Upsilon (2S) $
& 10023.26 $ \pm $ 0.31 & 9987 & 630 & 0.445 &
$ 0.416 \pm 0.004 \pm - \pm - $ \\
$ 3 \, {^{3} {\rm S}_{1}} $ & $ \Upsilon (3S) $
& 10355.2 $ \pm $ 0.5 & 10321 & 552 & 0.466 &
$ 0.433 \pm 0.003 \pm - \pm - $  \\
$ 4 \, {^{3} {\rm S}_{1}} $ & $ \Upsilon (4S) $
& 10579.4 $ \pm $ 1.2 & 10588 & 517 & 0.478 &
$ 0.493 \pm 0.003 \pm 0.013 \pm 0.002 $  \\
$ 5 \, {^{3} {\rm S}_{1}} $ & $ \Upsilon (10860) $
& 10865 $ \pm $ 8 & 10820 & 497 & 0.484 &
$ 0.424 \pm 0.003 \pm 0.078 \pm 0.011 $  \\
$ 6 \, {^{3} {\rm S}_{1}} $ & $ \Upsilon (11020) $
& 11019 $ \pm $ 8 & 11034 & 506 & 0.481 &
$ 0.508 \pm 0.003 \pm 0.057 \pm 0.012 $  \\
\hline
$
\begin{array}{c}
1 \, {^{3} {\rm P}_{2}} \\
1 \, {^{3} {\rm P}_{1}} \\
1 \, {^{3} {\rm P}_{0}}
\end{array}
$ &
$
\begin{array}{c}
\chi_{b2} (1P) \\
\chi_{b1} (1P) \\
\chi_{b0} (1P)
\end{array}
$ &
$
\left.
\begin{array}{c}
9912.21 \pm 0.26\pm 0.31 \\
9892.78 \pm 0.26\pm 0.31 \\
9859.44 \pm 0.42 \pm 0.31
\end{array}
\right\} 9899.87
$ & 9880 & 387 & 0.528 & $ 0.519 \pm 0.002 \pm - \pm - $  \\
$
\begin{array}{c}
2 \, {^{3} {\rm P}_{2}} \\
2 \, {^{3} {\rm P}_{1}} \\
2 \, {^{3} {\rm P}_{0}}
\end{array}
$ &
$
\begin{array}{c}
\chi_{b2} (2P) \\
\chi_{b1} (2P) \\
\chi_{b0} (2P)
\end{array}
$ &
$
\left.
\begin{array}{c}
10268.65 \pm 0.22 \pm 0.50 \\
10255.46 \pm 0.22 \pm 0.50 \\
10232.5 \pm 0.4 \pm 0.50
\end{array}
\right\}
10260.24
$ & 10231 & 343 & 0.549 & $ 0.524 \pm 0.002 \pm - \pm 0.001 $  \\
\end{tabular}
}
\end{table}

\newpage

\begin{table}[!ht] 
\footnotesize{
\caption{Light-heavy quarkonium systems
\vspace{0.3cm}}
\hspace{-2.truecm}\begin{tabular}{ccccccc}
  & & & & & &  \\
States & (MeV) & $m_{\rm exp}$ & $m_{\rm th}$ & $ Q $ & $
\alpha_{\rm E}^{(3)} $ &
$ \alpha_{\rm s}^{\rm exp}\pm\Delta_{\rm NLO}\pm\Delta_{\rm \Gamma}\pm
 \Delta_{\rm exp} $
\\
\hline
$ q \bar{c}$ &  &  &  &  &  &   \\
$ 1 \, {^{1} {\rm S}_{0}} $ &
$
\left\{
\begin{array}{c}
D^{\pm} \\
D^{0}
\end{array}
\right.
$
&
$
\left.
\begin{array}{c}
1869.3 \pm 0.4  \\
1864.5 \pm 0.4
\end{array}
\right\}
1867.7
$
& 1843 & 459 & 0.498   & $ 0.488 \pm 0.082 \pm - \pm - $  \\
$ 1 \, {^{3} {\rm S}_{1}} $ &
$
\left\{
\begin{array}{c}
D^{\ast}(2010)^{\pm} \\
D^{\ast}(2007)^{0}
\end{array}
\right.
$
&
$
\left.
\begin{array}{c}
2010.0 \pm 0.4  \\
2006.7 \pm 0.4
\end{array}
\right\}
2008.9
$
& 2000 &  &  & $ 0.499 \pm0.082 \pm - \pm - $  \\
$ 1 \Delta {\rm SS} $ &  & 141 $ \pm $ 1 & 157 &  &  &   \\
\hline
$
\begin{array}{c}
\\
1 \, {\rm P}_{2} \\  \\
1 \, {\rm P}_{1} \\  \\
\end{array}
$ &
$
\begin{array}{c}
\left\{
  \begin{array}{c}
    D_{2}^{\ast} (2460)^{\pm}  \\
    D_{2}^{\ast} (2460)^{0}
  \end{array}
\right.
\\
\left\{
  \begin{array}{c}
    D_{1} (2420)^{\pm}  \\
    D_{1} (2420)^{0}
  \end{array}
\right.
\\
\end{array}
$ &
$
\left.
\begin{array}{c}
  \begin{array}{c}
    2459 \pm 4   \\
    2461.1 \pm 1.6
  \end{array}
\\
  \begin{array}{c}
    2423.4 \pm 3.1   \\
    2422.3 \pm 1.3
  \end{array}
\\
\end{array}
\right\}
$ & 2443 & 232 & 0.619 & $ 0.651 \pm 0.051 \pm 0.028 \pm 0.005 $ \\
\hline
\hline \\[-3.5mm]
$ q \bar{b}$ &  &  &  &  &  &  \\
$ 1 \, {^{1} {\rm S}_{0}} $ &
$
\left\{
\begin{array}{c}
B^{\pm} \\
B^{0}
\end{array}
\right.
$
&
$
\left.
\begin{array}{c}
5279.0 \pm 0.5  \\
5279.4 \pm 0.5
\end{array}
\right\}
5279.1
$
& 5246 & 516 & 0.478 & $ 0.456 \pm 0.036 \pm - \pm - $  \\
$ 1 \, {^{3} {\rm S}_{1}} $ &
$ B^{\ast} $ & 5325.0 $ \pm $ 0.6 & 5311 &  &  &
$ 0.471 \pm 0.036 \pm - \pm - $  \\
$ 1 \Delta {\rm SS} $ &  & 46 $ \pm $ 3 & 64  &  &   &   \\
\end{tabular}
}
\end{table}

\begin{table} [!ht]
\vspace{-0.1truecm}
\footnotesize{
\caption{Light-heavy quarkonium systems
\vspace{0.3cm}}
\hspace{-2.truecm}\begin{tabular}{ccccccc}
  & & & & & & \\
States & (MeV) & $m_{\rm exp}$ & $m_{\rm th}$  & $ Q $ & $
\alpha_{\rm E}^{(3)} $ &
$ \alpha_{\rm s}^{\rm exp}\pm\Delta_{\rm NLO}\pm\Delta_{\rm \Gamma}\pm
 \Delta_{\rm exp} $
\\
\hline
$ s \bar{c} $ &  &  &  &   &  &  \\
$ 1 \, {^{1} {\rm S}_{0}} $ &
$ D_{s}^{\pm} $ & 1968.2 $ \pm $ 0.5 & 1959 & 472 & 0.493 &
$ 0.494 \pm 0.055 \pm - \pm - $  \\
$ 1 \, {^{3} {\rm S}_{1}} $ &
$ D_{s}^{\ast \pm} $ & 2112.0 $ \pm $ 0.6 & 2109 & &  &
$ 0.497 \pm 0.055 \pm - \pm - $  \\
$ 1 \Delta {\rm SS} $ &  & 144 $ \pm $ 1 & 149 &  &  &   \\
\hline
$
\begin{array}{c}
1 \, {\rm P}_{2} \\
1 \, {\rm P}_{1}
\end{array}
$ &
$
\begin{array}{c}
D_{s2} (2573)^{\pm} \\
D_{s1} (2536)^{\pm} \\
\end{array}
$ &
$
\left.
\begin{array}{c}
2573.5 \pm 1.7 \\
2535.35 \pm 0.34 \pm 0.5 \\
\end{array}
\right\}
$ & 2545 & 236 & 0.616 & $ 0.626 \pm 0.036 \pm 0.009 \pm 0.002 $ \\
\hline
\hline \\[-3.5mm]
$ s \bar{b} $ &  &  &  &  &  &  \\
$ 1 \, {^{1} {\rm S}_{0}} $ &
$ B_{s}^{0} $ & 5367.5 $ \pm $ 1.8 & 5343 & 535 & 0.472 &
$ 0.457 \pm 0.024 \pm - \pm 0.001 $ \\
$ 1 \, {^{3} {\rm S}_{1}} $ &
$ B_{s}^{\ast} $ & 5412.8 $ \pm $ 1.7 & 5408 &  &  &
$ 0.473 \pm 0.024 \pm - \pm 0.001 $  \\
$ 1 \Delta {\rm SS} $ &  & 47 $ \pm $ 4 & 65  &  &  &   \\
\hline $ 1 \, {\rm P} $ & $ B_{sJ}^{\ast} (5850) $ & 5853 $ \pm $
15 & 5830 & 255 & 0.602 & $ 0.592 \pm 0.013 \pm 0.042 \pm 0.027 $ \\
\end{tabular}
}
\end{table}


\newpage
\section{3-loop analytic coupling}

As mentioned, we compare our results with the 3-loop analytic
coupling~(\ref{ARC}), the difference with respect to the 4-loop
approximation being negligible~\cite{APT-07}. The exact 3-loop spectral
density~(\ref{RhoDef}) has a rather cumbersome structure in terms of the
Lambert function (see Ref.~\cite{Kurashev:2003pt}). However, for practical
purposes, it can be approximated with a high accuracy by the discontinuity
of the 3-loop perturbative running coupling (as given by PDG~\cite{pdg})
across the physical cut~\cite{Review}:
\be
\rho_1^{(3)}(\sigma)=\frac{1}{\beta_0}\frac{1}{(t^2+\pi^2)^3}
\left[t(3\pi^2-t^2)J(t)-(3t^2-\pi^2)R(t)\right].
\lb{sp3}
\ee
In this equation $t=\ln(\sigma/\Lambda^2)$,
\bea
J(t)&=&2t-B_1[tG_1(t)+G_2(t)]+B_1^2G_1(t)[2G_2(t)-1],\\
R(t)&=&t^2-\pi^2-B_1[tG_2(t)-\pi^2G_1(t)]+ \nonumber\\
&& B_1^2[G_2^2(t)-\pi^2G_1^2(t)-G_2(t)-1]+B_2,
\eea
\be
G_1(t)=\frac{1}{2}-\frac{1}{\pi}\arctan\!\left(\frac{t}{\pi}\right)\,,\quad
G_2(t)=\frac{1}{2}\ln(t^2+\pi^2)
\ee
with $B_j=\beta_j/\beta_0^{j+1}$ being the combination of the
$\beta$~function expansion coefficients
\begin{eqnarray}
\beta_{0} &=& \frac{1}{4 \pi} \left(11 - \frac{2}{3}\nf\right), \\
\beta_{1} &=& \frac{1}{(4 \pi)^2} \left(102 - \frac{38}{3}\nf\right), \\
\beta^{\overline{\rm MS}}_{2} &=& \frac{1}{(4 \pi)^3} \left(\frac{2857}{2} -
              \frac{5033}{18}\nf + \frac{325}{54}\nf^2\right).
\end{eqnarray}

In Fig.~\ref{ptan} the 3-loop analytic coupling~(\ref{ARC}) is compared to
the same level perturbative expression, both normalized at the Z~boson
mass to the world average value $\alpha_{\rm s}(M_Z^2) = 0.1176 \pm
0.0020\,$~\cite{pdg} and evolved at the heavy quark thresholds crossing by
continuous matching conditions. This gives for the analytic coupling the
scaling constant $\Lambda_{n_f=3}^{(3)} \simeq (417 \pm 42)\,$MeV in the
IR domain.


\clearpage
\newpage

\clearpage
\newpage

\clearpage
\newpage

\begin{figure}[th]
\centerline{\includegraphics[width=125mm]{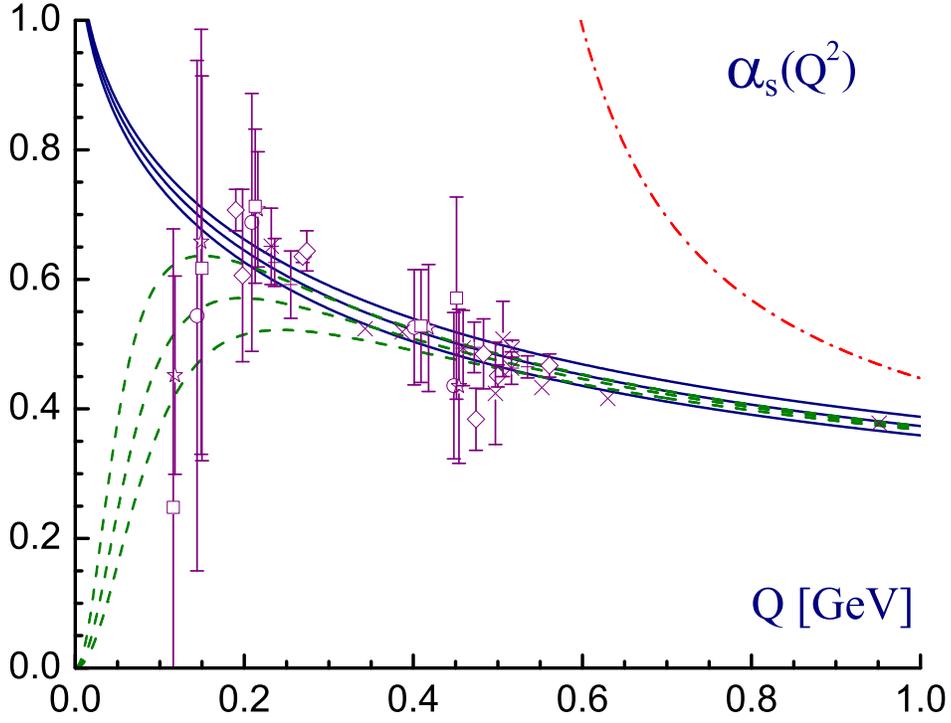}}
\vskip7.5mm
\caption{Comparison between points extracted from Tables I-VII in App.~B
and the three-loop analytic coupling~(\ref{ARC}) with
$\Lambda^{(3)}_{n_f=3} = (417 \pm 42)\,$MeV (solid curves). The
``massive'' one-loop analytic running coupling~(\ref{MARC}) (dashed
curves) refers to $\Lambda^{(1,{\rm eff})}_{n_f=3}=204\,$MeV and the
effective mass $m\ind{}{eff}=(38 \pm 10)\,$MeV. Three-loop perturbative
coupling (dot-dashed curve) corresponds to $\Lambda^{(3)}_{n_f=3} =
318\,$MeV. Circles, stars and squares refer respectively to $q\bar q\,$,
$s\bar s\,$ and $q\bar s\,$ with $q=u,d$ (light-light states). Heavy-heavy
states, $c\bar c\,$ and $b\bar b\,$, are represented by diamonds and
crosses. Finally in the light-heavy sector, asterisks stay for $q\bar c\,$
and $q\bar b\,$, whereas plus signs stand for $s\bar c\,$ and $s\bar b\,$.
Data for triplet and singlet states referring to the same multiplet are
combined in a weighted average according to their errors. Error bars
include both theoretical and experimental uncertainty and are drawn only
if relevant.}
\lb{low}
\end{figure}

\clearpage
\newpage

\begin{figure}[th]
\centerline{\includegraphics[width=125mm]{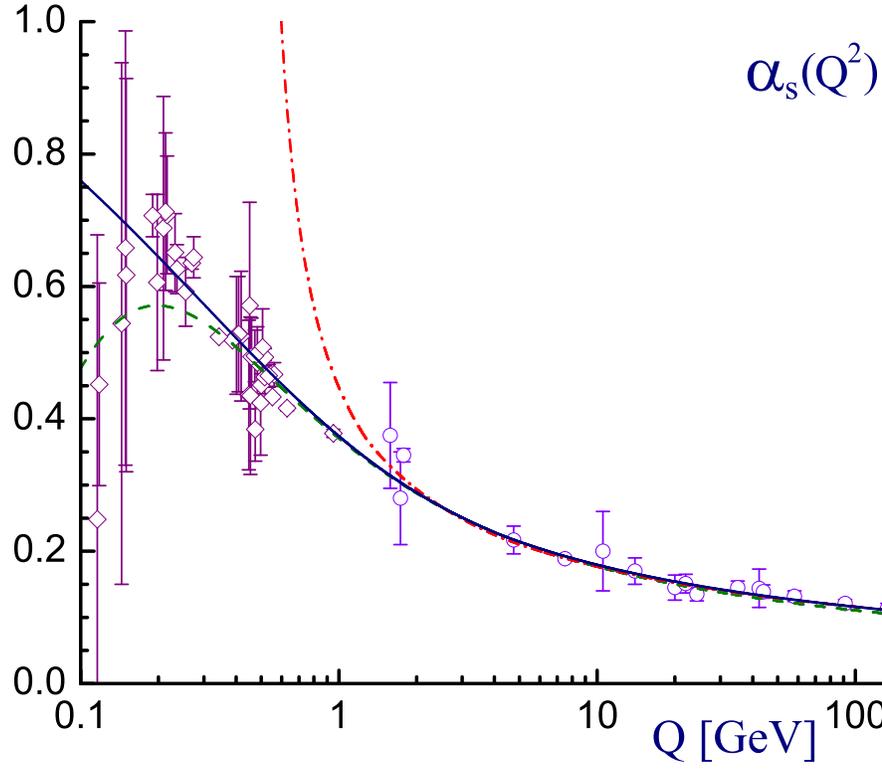}}
\vskip7.5mm
\caption{Summary of low ($\diamond$) and high energy ($\circ$) data
against the three-loop analytic coupling~(\ref{ARC}) (solid curve) and its
perturbative counterpart (dot-dashed curve) both normalized at the Z~boson
mass. Also shown is the ``massive'' one-loop analytic
coupling~(\ref{MARC}) (dashed curve) same as in Fig.~\ref{low}.}
\lb{tot}
\end{figure}

\clearpage
\newpage

\begin{figure}[th]
\centerline{\includegraphics[width=125mm]{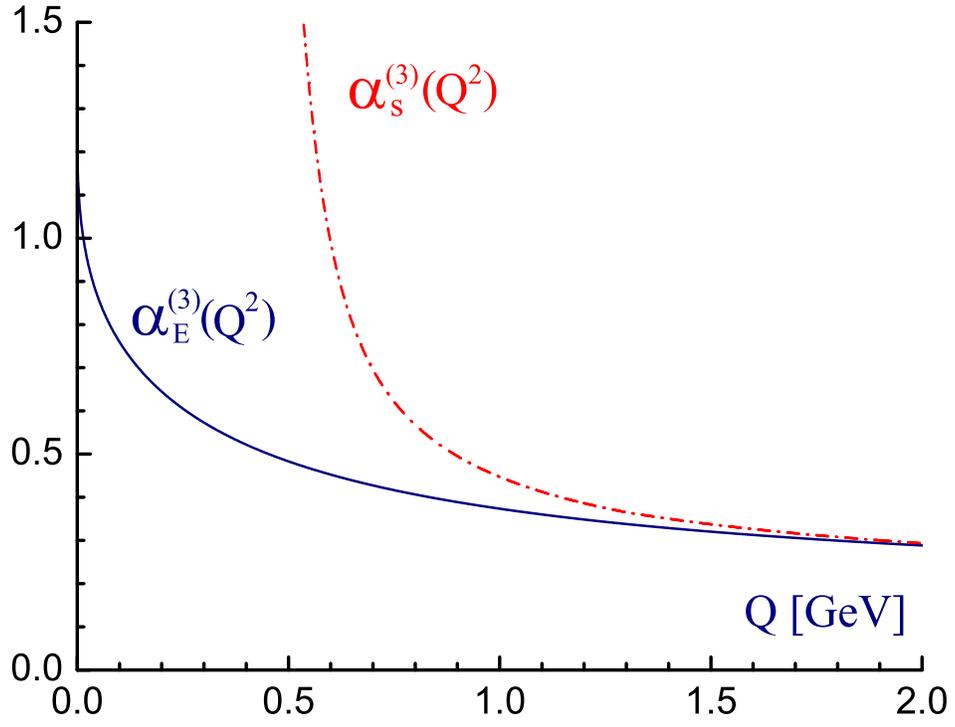}}
\vskip7.5mm
\caption{Three-loop QCD coupling below 2~GeV: the analytic
expression~(\ref{ARC}) (solid curve) versus the perturbative one
(dot-dashed curve), both normalized at the Z~boson mass according to the
world average~[31].}
\lb{ptan}
\end{figure}

\end{document}